\pdfoutput=1
\documentclass[sigconf, screen, nonacm]{acmart}

\ifdefined\pdfsuppressptexinfo \pdfsuppressptexinfo=-1 \fi

\usepackage{textcomp}
\usepackage{algorithm}
\usepackage{algpseudocode}
\usepackage{macros} % custom packages here


\begin{document}

\title[\textsc{ANOR}: Scaling Sparse Attention with a GPU--PNM Computing Hierarchy]
      {\sys: Scaling Sparse Attention \\ with a GPU–PNM Computing Hierarchy}

%% ---- Authors ----
\author{Hyunsu Ye}
\affiliation{%
  \institution{KAIST}
  \city{Daejeon}
  \country{Republic of Korea}}
\email{hsye@casys.kaist.ac.kr}

\author{Hyunmin Choi}
\affiliation{%
  \institution{KAIST}
  \city{Daejeon}
  \country{Republic of Korea}}
\email{hmchoi@casys.kaist.ac.kr}

\author{Jaehong Cho}
\affiliation{%
  \institution{KAIST}
  \city{Daejeon}
  \country{Republic of Korea}}
\email{jhcho@casys.kaist.ac.kr}

\author{Jaeseo Lee}
\affiliation{%
  \institution{KAIST}
  \city{Daejeon}
  \country{Republic of Korea}}
\email{stackoverflow@kaist.ac.kr}

\author{Hyeongseok Gwak}
\affiliation{%
  \institution{SK hynix}
  \country{Republic of Korea}}
\email{hyeongseok1.gwak@sk.com}

\author{Joohee Lee}
\affiliation{%
  \institution{SK hynix}
  \country{Republic of Korea}}
\email{joohee.lee@sk.com}

\author{Donguk Moon}
\affiliation{%
  \institution{SK hynix}
  \country{Republic of Korea}}
\email{donguk.moon@sk.com}

\author{Woosuk Chung}
\affiliation{%
  \institution{SK hynix}
  \country{Republic of Korea}}
\email{woosuk.chung@sk.com}

\author{Hoshik Kim}
\affiliation{%
  \institution{SK hynix}
  \country{Republic of Korea}}
\email{hoshik.kim@sk.com}

\author{Kiwan Maeng}
\affiliation{%
  \institution{The Pennsylvania State University}
  \city{University Park}
  \state{PA}
  \country{USA}}
\email{kvm6242@psu.edu}

\author{Jongse Park}
\affiliation{%
  \institution{KAIST}
  \city{Daejeon}
  \country{Republic of Korea}}
\email{jspark@casys.kaist.ac.kr}

\renewcommand{\shortauthors}{Ye et al.}

%% acmart requires the abstract before \maketitle
\begin{abstract}
Serving large language models with long contexts requires keeping the key-value (KV) cache of every request, which can exceed GPU memory capacity.
Sparse attention reads only a small, query-dependent subset of this cache at each decoding step, yet must retain the full cache because later steps may select other entries.
Processing-near-memory (PNM) devices attached through Compute Express Link (CXL) can hold the full cache and compute attention where it resides.
Prior GPU--PNM systems for sparse attention are PNM-centric, processing most of the attention on the PNM.
As existing CXL-PNM devices provide far less memory bandwidth than GPU HBM, this PNM-centric execution makes the PNM the bottleneck and leaves the GPU idle waiting for attention results.
We present \sys, a GPU--PNM joint acceleration system that exploits the \emph{selection locality} of sparse attention by computing most of the attention on the GPU from a cache of recently selected entries and processing only the misses on the PNM.
As the PNM processes cache misses directly, GPU cache updates can be decoupled from current-step execution.
We develop a bounded cache-update policy that limits updates to the point where the hit ratio of later steps saturates, reducing GPU-facing link traffic.
We add a small on-chip buffer, dubbed Early Streaming Buffer (ESB), that captures KV data during PNM attention and streams it to the GPU, avoiding most redundant DRAM reads despite the bandwidth mismatch between PNM memory and the GPU-facing link.
We further introduce an adaptive cache-factor mechanism that dynamically adjusts GPU cache size based on the serving load to balance cache hit ratio and batch size.
Across two models and three long-context agentic workloads, \sys improves decode throughput by 2.3$\times$ over HiSparse and 2.0$\times$ over a PNM baseline on average.
These results show that PNM is most effective not as the primary attention engine but as a computing tier of the GPU memory hierarchy.
\end{abstract}

\maketitle

%%%%%% -- PAPER CONTENT STARTS-- %%%%%%%%
\begin{figure}[t]
    \centering
    \includegraphics[width=0.9\columnwidth]{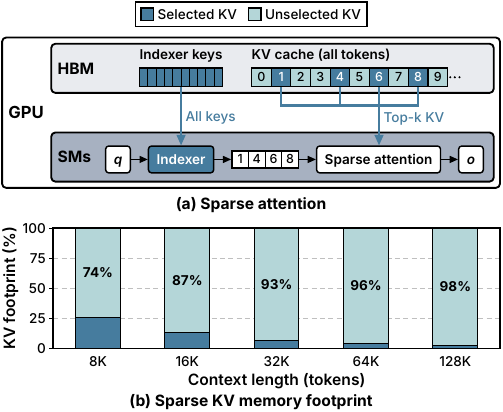}
    \vspace{-7pt}
    \caption{(a) Sparse attention via an indexer. (b) Share of the KV cache left unselected at each step on Qwen3-4B-Thinking with a Top-2K budget.}
    \label{fig:intro_inactive_kv}
    \vspace{-15pt}
\end{figure}

\section{Introduction}
\label{sec:intro}

% ===== P1: broad opening -> long context -> KV cache cost =====
Large language models (LLMs) increasingly serve long-context applications, ranging from chat assistants to coding agents and research tools~\cite{economic,sweagent,openhands,ai_scientist,chatgpt}.
Reasoning models generate long chains of intermediate tokens before answering~\cite{deepseek_r1,deepseek_v4,glm5}, while agentic workflows accumulate context across repeated model invocations and tool interactions~\cite{sweagent,react,reflexion}.
These growing contexts place substantial demands on both GPU memory bandwidth and capacity due to the expanding key-value (KV) cache.
The bandwidth demand slows down each decoding step, while the capacity demand limits the number of concurrent requests that fit in GPU memory, reducing batch sizes and serving throughput.

% ===== P2: sparse attention -> native trend -> capacity remains =====
Sparse attention addresses the per-step bandwidth demand.
Figure~\ref{fig:intro_inactive_kv}(a) illustrates how sparse attention reduces KV cache accesses during decoding.
At each decoding step, a lightweight indexer selects a small set of KV entries most relevant to the current query, allowing attention to read only the selected entries.
A growing number of models now perform sparse attention natively at inference time, either by incorporating it during pre-training~\cite{nsa,deepseek_v32,glm5,deepseek_v4,minimax_m3} or by applying it through post-training~\cite{seerattention_r,sas,infllmv2}.
However, sparse attention leaves the capacity demand unchanged.
As Figure~\ref{fig:intro_inactive_kv}(b) shows, at a 128K-token context, 98\% of the KV cache remains unselected at each step, yet the full cache must be retained because later steps may select different entries.

% ===== P3: KV offloading (locality, PCIe) -> PNM (PNM-centric, GPU idle) =====
KV offloading systems~\cite{hisparse,echo,arkvale,infllm,spin,pulseinfer,nosa} address this capacity demand by keeping the full KV cache in host memory and holding only recently selected entries in a GPU cache, exploiting the overlap of selections across consecutive decoding steps.
Yet, selected entries absent from the GPU cache, or \emph{misses}, must be fetched over the narrow PCIe link before attention, placing these transfers on the critical path of every decoding step.
To remove this bottleneck, several recent works~\cite{nelssa,longsight,scalablepnm,karat} compute sparse attention where the KV cache resides, using processing-near-memory (PNM) devices attached through Compute Express Link (CXL).
The GPU then exchanges only queries and attention results with the PNM, which processes most of the attention in a PNM-centric execution.
Such PNM devices, built on commodity DRAM, provide far less memory bandwidth than GPU HBM~\cite{marvell_structera_a,marvell_structera_sk_hynix_2026,xcena_mx1,h100nvl}.
As sparse attention reads only the selected entries, the GPU computes the same attention much faster than the PNM, making the PNM the bottleneck and leaving the GPU idle.

% ===== P4: \sys introduction and execution model =====
To this end, we propose \sys, a locality-aware GPU--PNM joint acceleration system for sparse attention.
\sys exploits selection locality to keep most of the attention on the GPU without holding the full KV cache in its memory or incurring the PCIe bottleneck of conventional KV offloading.
\sys partitions attention based on GPU cache residency: (1) the GPU computes attention over cached KV entries (\emph{hits}), while (2) the PNM, which holds the full KV cache, processes the remaining entries (\emph{misses}).
This allows the GPU to handle most of the attention as long as a high cache hit ratio is maintained.
Crucially, GPU cache updates are not required for current-step execution because the PNM processes the misses.
These updates affect only \emph{future locality}, allowing their timing and volume to be managed independently of current-step execution.

% ===== P5: challenges =====
Keeping most of the attention on the GPU across decoding steps requires addressing three challenges:

\begin{itemize}[labelindent=0.3em,nolistsep,leftmargin=1.0em]
  \item \textbf{Challenge 1: How much to update.}
To preserve a high hit ratio, the GPU cache must be updated continuously as selections change, and the updates travel over the PCIe link between the GPU and the PNM devices (the \emph{GPU-facing link}).
Updating every miss saturates this link, whereas updating too few lowers the hit ratio and shifts attention back to the PNM, leaving the GPU idle.

  \item \textbf{Challenge 2: Bandwidth mismatch between PNM memory and the link.}
The GPU-facing link is much slower than the internal memory bandwidth of the PNM, so the update data cannot be sent to the GPU at the rate the PNM reads it for attention.
The PNM must instead read the same data again for cache updates, which increases DRAM traffic and may slow its attention.

  \item \textbf{Challenge 3: Load-dependent cache size.}
A larger GPU cache reduces the load on the PNM but leaves less HBM for concurrent requests, whereas a smaller one admits more requests but shifts more attention to the PNM.
The best balance shifts as the serving load changes, so a fixed cache size either limits the batch under high load or forgoes hit ratio under low load.
\end{itemize}

% ===== P6: transition to design mechanisms =====
To address these challenges, \sys takes a hardware--runtime co-design approach that preserves the organization of existing CXL-PNM devices while adding only lightweight hardware support.
Our design consists of three complementary mechanisms, each addressing one of the three challenges described above:

\begin{description}[leftmargin=1.0em]

\item \textbf{(1) Bounded cache updates for future locality.}
\sys moves GPU cache updates off the current step.
The GPU and the PNM compute each step with the current cache contents, and updates are applied in the background, taking effect only in later steps.
As updates are not required for the correctness of the current step, \sys can choose how much to update instead of transferring every miss.
By characterizing the selections of sparse attention, we find that the hit ratio of later steps saturates once a certain fraction of the misses is updated.
\sys therefore updates only that fraction of the misses.
Updating more would consume link bandwidth for only marginal gains in hit ratio, so \sys keeps link traffic low while preserving the hit ratio.

\item \textbf{(2) Early Streaming Buffer (ESB).}
The PNM already reads the update data from memory when it computes attention over the misses.
The ESB exploits this by capturing the update data \emph{early}, as the PNM reads it for attention, rather than after attention completes.
Lightweight capture logic copies the data selected for update from the memory read path into an on-chip buffer, which drains to the GPU at the rate the link allows.
The buffer thus absorbs the bandwidth mismatch between PNM memory and the GPU-facing link, so most update data reaches the GPU directly from the reads for attention.
Because the update volume is bounded, a small buffer suffices.

\item \textbf{(3) Adaptive cache factor.}
The \emph{cache factor} is the GPU cache size of each request relative to the number of entries selected per step.
A larger cache factor raises the hit ratio but takes HBM away from concurrent requests, and the best choice depends on how many requests are waiting.
At runtime, \sys first decides how many requests to batch and then sets the cache factor calibrated offline for that batch size, lowering it only if HBM runs short.
Under low load, \sys keeps a large cache factor to maximize the hit ratio; under high load, it lowers the cache factor to admit more requests and restores it once the pending requests drain.

\end{description}

% ===== P7: evaluation =====
We evaluate \sys on a GPU system with CXL-PNM devices using (1) two models: Qwen3-4B-Thinking and Qwen3-30B-A3B-Thinking, and (2) three long-context agentic workloads: SWE-chat, Open-SWE, and OpenResearcher.
We combine measured GPU kernel latencies with a cycle-level CXL-PNM simulator~\cite{m2ndp} configured with publicly available device specifications~\cite{marvell_structera_sk_hynix_2026}.
\sys achieves 2.3$\times$ higher decode throughput than HiSparse and 2.0$\times$ higher than a PNM baseline, averaged over both models and three agentic workloads.
It also outperforms the best static GPU--PNM partition, which assigns a fixed share of attention to the GPU, by 1.43$\times$.
In online serving, \sys reduces P99 completion latency by 2.8$\times$ and 2.2$\times$ relative to the best-performing baseline.

% ===== P8: insights (replaces the contribution list) =====
These results show that \sys turns the PNM from a bottleneck into a computing tier behind the GPU.
By keeping most of the attention on the GPU and assigning only the misses to the PNM, \sys makes PNM a practical extension of GPU memory, enabling scalable long-context LLM serving beyond GPU memory capacity.
\section{Background}
\label{sec:background}

\subsection{Sparse Attention for Long-Context Decoding}
\label{sec:background:sparse_attention}

\noindent\textbf{Memory demand of long-context decoding.}
%
% After processing the prompt in a prefill phase, a large language model (LLM) generates its output one token at a time, and each of these decoding steps attends to all of the tokens that precede it.
A large language model (LLM) first processes the input prompt in a prefill phase and then enters a decoding phase, where it generates one token at a time by attending to all preceding tokens.
To avoid recomputing this history, the model stores the key and value vectors of every preceding token in a \emph{KV cache} and reads the whole cache at each step~\cite{pope2023scaling,vllm}.
%
% Therefore, the KV cache grows with its context length, and the KV caches of all requests served together in a batch, such as Req A and Req B in the figure, share the high-bandwidth memory (HBM) of the GPU.
This KV cache resides in the GPU's high-bandwidth memory (HBM) and grows with the request's context length.
The KV caches of all requests in a batch share this memory.
%
% Contexts have grown considerably longer, because recent reasoning models produce long chains of intermediate tokens~\cite{deepseek_r1,sparsespec} and agentic workflows accumulate context across model invocations and tool interactions~\cite{sweagent,infercept}.
Recently, contexts have grown considerably longer, because reasoning models produce long chains of intermediate tokens~\cite{deepseek_r1,sparsespec} and agentic workflows accumulate context across model invocations and tool interactions~\cite{sweagent,infercept}.
%
% As contexts lengthen, the KV cache demands more HBM bandwidth at every decoding step and more HBM capacity for every request in the batch~\cite{attacc,neupims,infinigen}.
%
As a result, the KV cache requires more HBM bandwidth at each decoding step and more HBM capacity per request in the batch~\cite{attacc,neupims}.

\begin{figure}[t]
    \centering
    \includegraphics[width=\columnwidth]{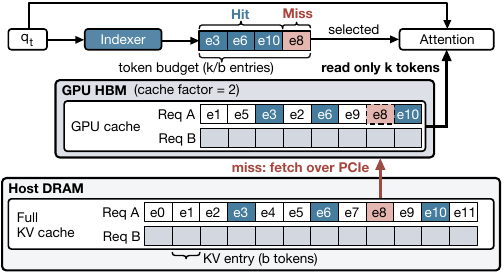}
    \caption{Sparse attention with a GPU cache. Of the entries that the indexer selects, hits are read directly from HBM, whereas a miss is first fetched from host DRAM over PCIe.}
    \label{fig:background_sparse}
    \vspace{-10pt}
\end{figure}

\noindent\textbf{Query-dependent sparse attention.}
To reduce the per-step bandwidth demand, sparse attention has been proposed~\cite{nsa,deepseek_v32,glm5}.
This approach restricts each query's attention to tokens with the highest expected attention scores.
A lightweight scoring component, called the \emph{indexer}, estimates these scores from compact metadata of the KV cache.
%
% The unit of selection is a \emph{KV entry}, which covers either a single token or a fixed number of consecutive tokens depending on the granularity.
Based on these scores, the indexer selects the KV cache in units of \emph{KV entries}.
%
% The total number of tokens covered by the selected entries is the \emph{token budget}~\cite{quest}.
The \emph{token budget} specifies the total number of tokens covered by the selected entries at each decoding step~\cite{quest}.
%
% For example, with $b$ tokens per entry and a token budget of $k$, the indexer selects $k/b$ entries, as shown in Figure~\ref{fig:background_sparse}(b). and each step reads the KV of only $k$ tokens instead of all $L$.
As shown in Figure~\ref{fig:background_sparse}, with $b$ tokens per entry and a token budget of $k$, the indexer selects $k/b$ entries, so each step reads the KV of $k$ tokens instead of all $L$ tokens in the context.
Depending on how the indexer is obtained, sparse attention designs are either training-free or training-based.
Training-free designs approximate the scores from the keys of an unmodified model at inference time~\cite{quest,infinigen,arkvale,retroinfer}, whereas training-based designs train the indexer, or the model with sparse selection, during pre-training or post-training so that sparse selection preserves accuracy~\cite{deepseek_v32,nsa,seerattention_r,infllmv2,sas}.
%
% TODO: undecided whether to keep the training-free vs. learned indexer contrast and the scope sentence below.
% Training-free designs such as Quest derive these scores from summaries of the keys~\cite{quest}, whereas learned designs such as DeepSeek-V3.2, SeerAttention-R, and SAS train the indexer to rank the context for the model~\cite{deepseek_v32,seerattention_r,sas}.
%
% This paper focuses on learned indexers that select a fixed token budget at each step.
%

\noindent\textbf{Retaining the full KV cache.}
Sparse attention reduces the data that each step reads, but it does not allow the system to discard the rest of the KV cache.
%
% Because the selection depends on the query, an entry that one step skips may be selected by a later step~\cite{quest}.
This is because KV entries not selected in the current step may be selected in later steps~\cite{quest}.
The system must therefore keep the full KV cache of every request, even though each step reads only $k$ of its $L$ tokens.
%
% As a result, sparse attention relieves the per-step bandwidth demand but leaves the capacity demand in place, since the full KV cache still grows with the context length and the batch size.
As a result, sparse attention reduces per-step bandwidth demand but does not reduce the KV cache capacity required for a given context length and batch size.

\subsection{KV Cache Offloading with a GPU Cache}
\label{sec:background:gpu_offloading}

\noindent\textbf{GPU cache over host memory.}
KV cache offloading has been widely used to relieve the capacity pressure of the KV cache, moving it to a larger but slower memory, such as host DRAM, and transferring the needed parts to the GPU during decoding~\cite{flexgen,vllm,infinigen,mooncake}.
This approach also suits sparse attention well, because each step needs to transfer only its selected entries.
%
% Recent systems for sparse attention therefore manage the KV cache across a two-level memory hierarchy, keeping the full KV cache in host DRAM and only a bounded subset of KV entries in a \emph{GPU cache} in HBM~\cite{hisparse,echo}, as Figure~\ref{fig:background_sparse}(b) shows.
As shown in Figure~\ref{fig:background_sparse}, recent systems for sparse attention manage the KV cache across a two-level memory hierarchy, keeping the full KV cache in host DRAM and only a bounded subset of KV entries in a \emph{GPU cache} in HBM~\cite{hisparse,echo}.
At each step, the indexer on the GPU selects entries without reading the full KV cache.
Selected entries that reside in the GPU cache are read directly from HBM, whereas the cache misses are fetched from host memory before attention execution and then inserted into the GPU cache for later steps.
In this example, three of the four selected entries hit in the GPU cache, whereas the fourth misses and is fetched from host DRAM over PCIe.

\noindent\textbf{Selection locality across decoding steps.}
The benefit of this hierarchy depends on how many of the selected entries hit in the GPU cache.
%
% Consecutive decoding steps of the same request tend to select many of the same entries, so an entry brought into the GPU cache for one step is often selected again in later steps~\cite{hisparse,echo}.
Consecutive decoding steps of the same request tend to select many of the same entries~\cite{hisparse,echo}.
Sparse selections thus exhibit \emph{selection locality}, a form of temporal locality in KV accesses~\cite{hisparse}.
As a result, a GPU cache much smaller than the full KV cache can still serve many of the entries selected at each decoding step.

\noindent\textbf{Cache factor and batch size.}
For each request and layer, we define the \emph{cache factor} as the number of tokens the GPU cache can hold divided by the token budget.
The GPU cache in Figure~\ref{fig:background_sparse}, for example, holds eight entries for a token budget of four entries, a cache factor of two.
A larger cache factor keeps more of the recently selected entries and therefore serves more of each selection from HBM.
However, because the GPU cache of every request occupies HBM, a larger cache factor also limits the batch size.

\subsection{GPU--PNM Systems for Sparse Attention}
\label{sec:background:cxl_pnm}
\begin{figure}[t]
    \centering
    \includegraphics[width=0.83\columnwidth]{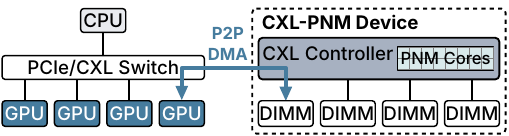}
    \vspace{-7pt}
    \caption{A GPU system with a CXL-PNM device.}
    \label{fig:background_gpu_pnm}
    \vspace{-13pt}
\end{figure}

\noindent\textbf{CXL memory expansion and processing near memory.}
Compute Express Link (CXL) attaches additional memory devices to a host over an interconnect built on the PCIe physical layer~\cite{pond,tpp,m2ndp}.
CXL memory expanders therefore extend memory capacity beyond both HBM and host DRAM, allowing the system to hold the full KV caches of many long-context requests.
Some expanders further integrate compute units next to their DRAM, an approach known as \emph{processing near memory} (PNM).
Memory and chip vendors have released such CXL-PNM devices as products and prototypes~\cite{cms,cxlpnm,marvell_structera_a,marvell_structera_sk_hynix_2026,xcena_mx1}.
Because these PNM units read KV data directly from the DRAM channels inside the expander, they can exploit the high aggregate bandwidth of these channels~\cite{m2ndp}.
As shown in Figure~\ref{fig:background_gpu_pnm}, the device can also exchange data directly with the GPU through peer-to-peer (P2P) DMA over PCIe.

\noindent\textbf{Attention across the GPU and PNM.}
Once a request's KV cache resides on a PNM device, sparse attention can run entirely on the PNM cores, so that only queries, entry indices, and attention results need to be exchanged between the GPU and the device.
Sparse attention can also be split between the GPU and the PNM device without changing its result.
Each device computes attention over its own subset of the selected entries and returns a partial output together with the log-sum-exp (LSE) of its attention scores.
These partial outputs are then rescaled using their LSE values and summed to recover the same result as attention over the complete set.
%

% \niparagraph{Commercial CXL-PNM.}
% %
% Current CXL-PNM products share a common structure~\cite{marvell_structera_a,marvell_structera_sk_hynix_2026,xcena_mx1}.
% %
% Each device integrates general-purpose cores and DDR5 memory controllers into a CXL controller chip and attaches commodity DDR5 DRAM to it.
% %
% Marvell Structera A, for example, integrates 16 Arm Neoverse V2 cores and four DDR5-6400 channels, which provide about 200~GB/s and up to 4~TB~\cite{marvell_structera_a,nelssa}.
% %
% This structure suits sparse attention over long contexts.
% %
% Commodity DDR5 DIMMs give each device terabytes of capacity, enough to hold the full KV caches of many long requests.
% %
% Adding devices also raises the aggregate DRAM bandwidth available to the PNM cores.
% %

% \niparagraph{Our target.}
% %
% This work targets GPU systems with DDR5-based CXL-PNM devices, so its design builds naturally on top of commercial hardware and remains practical to deploy.
%
\niparagraph{CXL-PNM device organization.}
Current CXL-PNM devices share a common organization~\cite{marvell_structera_a,marvell_structera_sk_hynix_2026,xcena_mx1}: general-purpose cores and memory controllers are integrated into the CXL controller chip, which attaches commodity DRAM modules.
Marvell Structera A, for example, integrates 16 Arm Neoverse V2 cores with four DDR5-6400 channels, providing about 200~GB/s and up to 4~TB per device~\cite{marvell_structera_a,nelssa}.
This organization suits sparse attention for long-context workloads, because commodity DIMMs provide terabytes of capacity per device, enough to hold the full KV caches of \mbox{many long requests.}

\section{Motivation}
\label{sec:motivation}

This section first shows that PNM-centric execution leaves the GPU underutilized (\S\ref{sec:motivation_commercial}), and then examines three challenges that arise when the GPU instead serves most of the attention from a cache of selected entries (\S\ref{sec:motivation_refill}--\S\ref{sec:motivation_adaptive}).
Unless otherwise noted, the experiments in \S\ref{sec:motivation} and \S\ref{sec:principles} model a GPU with CXL-PNM devices using the hardware configuration in \S\ref{sec:methodology}.
They use Qwen3-4B-Thinking with a Top-2K budget, replay the selection traces of three trained indexers over 512 consecutive decoding steps on three agentic workloads, and report the average over indexers.

\begin{figure}[t]
    \centering
    \includegraphics[width=0.95\linewidth]{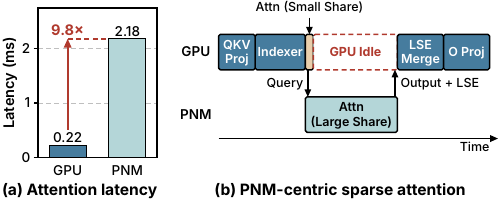}
    \caption{Attention execution imbalance between the GPU and
     CXL-PNM under PNM-centric partitioning.
     % \todo{}{State the model and the setting of this measurement in the figure caption.=> (hsye) Qwen3-4B-thinking with top-2k sparse attention, 1 layer, batch 64, M2NDP tuned based on Marvell CXL-PNM performance}
     }
    \label{fig:motivation_partition}
    \vspace{-5pt}
\end{figure}

\begin{figure}[t]
    \centering
    \includegraphics[width=0.95\linewidth]{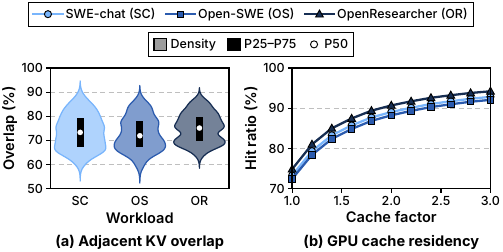}    
    \caption{Selection locality of sparse attention. (a) Overlap between the KV entries selected at adjacent decoding steps. (b) GPU cache hit ratio over increasing cache factor.}
    \label{fig:selection_locality}
    \vspace{-10pt}
\end{figure}

\subsection{GPU Underutilization in PNM-Centric Sparse Attention}
\label{sec:motivation_commercial}
% \todo{hmchoi}{rewrite paragraph 1}
Prior GPU--PNM systems run most of sparse attention on the PNM devices and leave only a small or fixed share, such as recent tokens, to the GPU~\cite{karat,longsight,scalablepnm,nelssa}.
This PNM-centric execution does not fit existing CXL-PNM devices, which provide terabytes of memory capacity but far less memory bandwidth than GPU HBM~\cite{marvell_structera_a,marvell_structera_sk_hynix_2026,xcena_mx1,h100nvl}.
When the GPU and the PNM devices compute the same sparse attention for a batch of 64 requests at a single layer, the PNM devices take nearly ten times longer, reflecting this bandwidth gap (Figure~\ref{fig:motivation_partition}(a)).
Under PNM-centric execution, the GPU thus finishes its small share quickly and waits idle while the PNM processes the large share (Figure~\ref{fig:motivation_partition}(b)), which lowers the throughput of the whole system.
Minimizing this idle time requires assigning as much of the attention as possible to the GPU.

The GPU can take most of the attention without holding the full KV cache, because consecutive decoding steps select many of the same entries, so a small GPU cache can serve most of each selection.
Recent offloading systems exploit this \emph{selection locality} to bridge the gap between the full KV cache and HBM capacity: they keep recently selected entries in a GPU cache and fetch only the misses from host memory~\cite{hisparse,echo,arkvale,infllm,spin,pulseinfer,nosa} (\S\ref{sec:background:gpu_offloading}).
Figure~\ref{fig:selection_locality} measures this selection locality on three agentic workloads, SWE-chat, Open-SWE, and OpenResearcher~\cite{swechat,openswe,openresearcher}.
A GPU cache with a cache factor of one already serves more than 70\% of the selected entries, and a cache factor of three raises the hit ratio above 90\% on all three workloads.
In a GPU--PNM system, the GPU computes attention over the cache hits (\emph{hit attention}), while the PNM computes attention over the misses in place (\emph{miss attention}), so misses need not reach the GPU before the current attention.
Exploiting selection locality in this way, however, raises three challenges, which the following subsections examine.

\begin{figure}[t]
    \centering
    \includegraphics[width=0.9\linewidth]{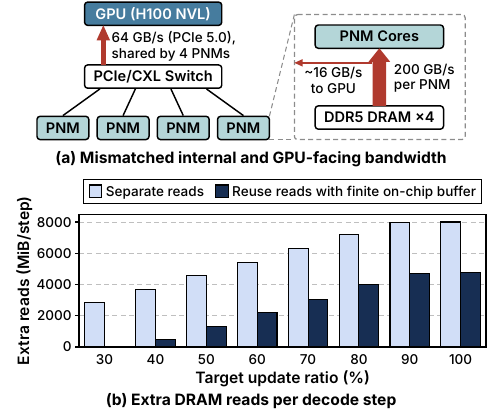}
    \caption{(a) Internal and GPU-facing bandwidth of a CXL-PNM device. (b) Extra DRAM reads for cache updates.}
    \label{fig:design_principle_read_reuse}
    \vspace{-10pt}
\end{figure}

\subsection{Challenge 1: Updating the GPU Cache over a Narrow GPU-Facing Link}
\label{sec:motivation_refill}

Preserving selection locality requires continuous GPU cache updates, and this update traffic remains even when the PNM computes the current misses.
Because sparse selections change across decoding steps, previously missing entries must be brought into the GPU cache as they become useful for future steps.
Computing current-step misses on the PNM removes these transfers from the current attention dependency, but it does not eliminate the cache-update traffic required to preserve future GPU locality.

This traffic is problematic because the GPU-facing link is narrow, as it is shared by the PNM devices, and transferring every miss over PCIe is a well-known bottleneck in offloading systems~\cite{hisparse,echo}.
Although updates lie off the critical path of the current step, each update must still complete before the same layer executes in the next decoding step.
Updating every miss sends more data than the link can deliver before this deadline, so the GPU again waits for transfers, reintroducing the PCIe bottleneck of KV offloading.
Conversely, updating too few misses lowers the GPU cache hit ratio, increasing the fraction of future attention processed by the slower PNM, which again leaves the GPU waiting for its results.

\textbf{Challenge 1:} \textit{Cache updates are needed to preserve GPU locality, yet updating every miss reintroduces the PCIe bottleneck before the next step, while updating too few lowers the hit ratio.}

\subsection{Challenge 2: Bandwidth Mismatch between PNM Memory and the GPU-Facing Link}
\label{sec:motivation_reuse}

Cache updates also burden the PNM, because each device reads its memory far faster than it can send data to the GPU.
As Figure~\ref{fig:design_principle_read_reuse}(a) shows, each device reads its DRAM at about 200~GB/s, whereas its share of the GPU-facing link is only about 16~GB/s.
The PNM already reads the entries selected for GPU cache updates to compute attention over the misses, but it cannot send them to the GPU at this read rate without throttling its attention to the link rate.
A straightforward implementation therefore issues separate DRAM reads for these entries; because sparse attention on the PNM is memory-bound and already makes heavy use of local memory bandwidth, these reads may contend with it.

\textbf{Challenge 2:} \textit{Because the GPU-facing link is far slower than PNM memory, update data cannot be sent as the PNM reads it for attention, so cache updates require redundant DRAM reads of already-read data.}

\begin{figure}[t]
    \centering
    \includegraphics[width=\linewidth]{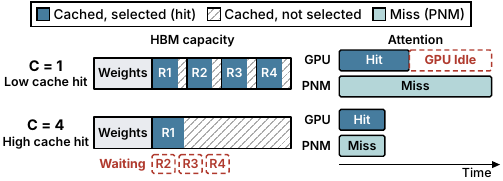}
    \vspace{-5pt}
    \caption{Cache factor trade-off.}
    \label{fig:cache_factor_tradeoff}
    \vspace{-10pt}
\end{figure}

\subsection{Challenge 3: Cache Factor Trade-off between PNM Load and Batch Size}
\label{sec:motivation_adaptive}
The cache factor determines how the GPU cache and request batching share HBM, as Figure~\ref{fig:cache_factor_tradeoff} illustrates.
A small cache factor leaves room for a larger batch, but more selections miss in the GPU cache, so the GPU waits longer for PNM attention.
A large cache factor raises the hit ratio and reduces PNM work, but the GPU cache also holds many entries that the current step does not select, which limits the batch size.
The best balance shifts with the number of active requests, which changes continuously in online serving.

\textbf{Challenge 3:} \textit{The cache factor trades PNM load against batch size, and the best balance shifts with concurrency during online serving.}
\section{Design Principles}
\label{sec:principles}
Each challenge in \S\ref{sec:motivation} stresses a different resource that the GPU cache depends on: the GPU-facing link (Challenge~1), PNM memory bandwidth (Challenge~2), and HBM capacity (Challenge~3).
We derive one principle for each.

\subsection{Principle 1: Bound Cache Updates at the Saturation Point of Future Locality}
\label{sec:principle_selective_update}

\begin{figure}[t]
    \centering
    \includegraphics[width=\linewidth]{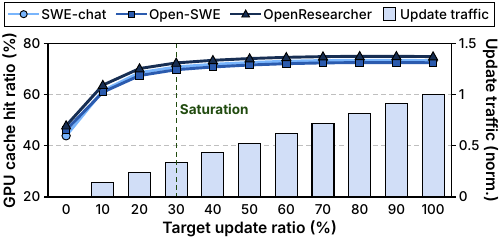}
    \caption{GPU cache hit ratio and normalized update traffic across target update ratios.}
    \label{fig:design_principle_selective_update}
    \vspace{-15pt}
\end{figure}

Challenge~1 (\S\ref{sec:motivation_refill}) leaves the system to decide how much of the miss traffic to send to the GPU.
This decision is possible because current-step execution does not depend on cache updates: the PNM computes attention over current-step misses, so updates serve only future steps.
The update amount thus becomes a parameter that trades link traffic for future hit ratio, rather than a requirement set by the current attention.
Figure~\ref{fig:design_principle_selective_update} shows the future GPU cache hit ratio, measured as the cache state evolves across consecutive decoding steps, and the update traffic, normalized to updating every miss, as functions of the target update ratio.
The hit ratio rises quickly and saturates by about a 30\% update ratio on all three workloads.
We measure this at cache factor 1.0, where misses are most frequent; larger cache factors reduce the miss volume and saturate no later.
The saturation also holds for each indexer and each workload independently: updating 30\% of the misses in recency order reaches 95.0--98.5\% of the hit ratio obtained by updating every miss for all three indexers on each of the three workloads, at about one third of the update traffic.
Updates beyond this point thus add link traffic rather than improving future locality, which \S\ref{sec:evaluation_sensitivity} also confirms end to end.

\textbf{Design implication.}
GPU cache updates should be decoupled from current-step execution and bounded to a budget near the saturation point rather than covering every miss (\S\ref{sec:system_selective_update}).

\begin{figure*}[t]
    \centering
    \includegraphics[width=0.95\textwidth]{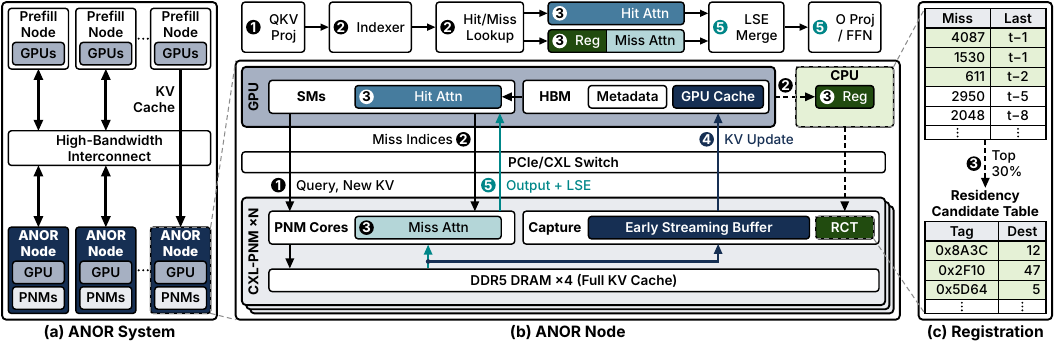}
    \vspace{-5pt}
    \caption{Overview of \sys.}
    \label{fig:system_overview}
    \vspace{-5pt}
\end{figure*}
\subsection{Principle 2: Reuse PNM Attention Reads to Bridge the Bandwidth Mismatch}
\label{sec:principle_data_reuse}

The key observation is that update candidates are a subset of the misses that the PNM already reads to compute attention, so their data passes through the PNM memory path during its attention.
Capturing this data as it is read and holding it until the link can send it avoids most separate reads without throttling PNM attention.
The required storage is small, because Principle~1 (\S\ref{sec:principle_selective_update}) bounds the update volume to a fraction of the misses.
Holding this data in the existing cache hierarchy is not reliable, because hardware-managed replacement may evict it before the link can send it.
Figure~\ref{fig:design_principle_read_reuse}(b) confirms this: with separate reads, extra DRAM reads grow with the update ratio, whereas reusing reads with a finite on-chip buffer removes nearly all of them at a 30\% update ratio.
At higher update ratios, the buffer overflows and some data must be read again, which further supports bounding the update amount.

\textbf{Design implication.}
GPU cache updates should capture their data from the PNM attention read stream into dedicated on-chip storage and drain it asynchronously to the GPU, avoiding most additional DRAM reads without stalling PNM attention (\S\ref{sec:system_streaming}).

\subsection{Principle 3: Match GPU Cache Allocation to Runtime Resource Availability}
\label{sec:principle_adaptive}

Challenge~3 (\S\ref{sec:motivation_adaptive}) implies that no single cache factor is best at all times, because the right balance depends on how much HBM the current load leaves free.
Under low load, the system should raise the cache factor to use spare HBM, so that the GPU serves more of each selection.
Under high load, the system should lower the cache factor to hold fewer unselected entries in HBM and admit more requests into the batch, raising the reuse of each weight read across requests.

\textbf{Design implication.}
The cache factor should be a runtime decision that follows the available HBM and the load, letting the GPU serve as much of each selection as the current load \mbox{permits (\S\ref{sec:system_adaptive}).}

\section{\sys Overview}
\label{sec:system_overview}

\sys applies the three principles of \S\ref{sec:principles} to a GPU with CXL-PNM devices.
Following prior disaggregated LLM serving systems~\cite{distserve,splitwise}, we separate prefill and decode and focus on the decode node, where long-context KV capacity and sparse-attention bandwidth become the primary bottlenecks.
We partition each request's KV cache across CXL-PNM devices by attention head, distributing miss attention across devices~\cite{nelssa}.
Figure~\ref{fig:system_overview} summarizes the decode-node organization and its GPU--PNM dataflow, including the asynchronous GPU cache update path.

\subsection{Execution Flow}
\label{sec:system_flow}

Each decoding step partitions the selected sparse KV entries according to the current GPU cache state so that the GPU can process as much sparse attention as its resident KV permits.
Figure~\ref{fig:system_overview} illustrates the five major stages of this execution flow.

\noindent\textbf{(1) Projection and KV propagation.}
The GPU performs QKV projection for the current layer, forwards the query to the PNM devices, and appends the newly generated KV to the PNM-resident full KV cache.

\noindent\textbf{(2) Sparse selection and hit/miss classification.}
The indexer produces the selected KV set on the GPU.
After sparse selection, GPU-side cache metadata lookup classifies the selected KV entries into GPU cache hits and misses.
The indices required for miss attention are forwarded to the corresponding PNM devices.

\noindent\textbf{(3) Parallel hit/miss attention and update registration.}
The GPU begins attention over its cache hits.
All cache misses are assigned to PNM attention; in parallel with GPU hit attention, the host selects only a bounded subset of these misses as \emph{update candidates} and registers them with the PNM devices.
After candidate registration, each PNM begins attention over its assigned cache misses.
This registration is lightweight, accounting for about 3.8\% of decoding latency.

\noindent\textbf{(4) Read-reusing cache update.}
While the PNM reads miss KV from local memory for attention, update candidates are captured from the same read stream and streamed toward the GPU (\S\ref{sec:system_streaming}).

\noindent\textbf{(5) Exact merge and subsequent computation.}
Once the GPU-hit and PNM-miss attention results are available, the GPU combines the partial outputs and their log-sum-exp statistics.
The GPU then proceeds with the output projection and FFN.
GPU cache updates continue in the background (\S\ref{sec:system_streaming}).

\noindent\textbf{Exactness.}
Our execution partition does not change the sparse-attention selection itself.
The same selected KV set is partitioned into GPU cache hits and PNM misses, and each side computes attention over its assigned subset.
The GPU then combines the partial attention outputs and their log-sum-exp statistics to recover the same result as executing the original sparse attention over the complete selected set.
GPU cache updates affect only where a selected KV entry may be processed in future steps; they do not change which KV entries participate in attention.
\subsection{Bounded Cache Updates for Future Locality}
\label{sec:system_selective_update}

Following Principle~1 (\S\ref{sec:principle_selective_update}), GPU cache updates serve only future locality because the PNM 
already handles current-step misses.

\noindent\textbf{Update ratio.}
For each layer and decoding step, the host assigns a bounded update ratio over the current GPU cache misses.
Following the saturation behavior in \S\ref{sec:principle_selective_update}, our default configuration updates 30\% of the current misses.
This ratio is defined over the miss set of the current selection, not the entire GPU cache.

\noindent\textbf{Candidate ordering.}
Among these misses, the host prioritizes candidates by recency.
Recency is determined by the most recent prior decoding step in which a KV entry appeared in the sparse selection for the same request, layer, and head.
This policy directly exploits selection locality and requires only the history of sparse selections, without relying on indexer-specific scores.

\noindent\textbf{Placement and readiness.}
Before an update is issued, the runtime reserves a destination entry in the GPU cache.
Eviction victims are chosen outside the current selected set so that cache maintenance does not evict KV entries being consumed by the current attention operation.
An updated entry becomes usable as a cache hit only after all required KV fragments have arrived at the GPU.
Cache updates proceed in the background and need to complete only before the same layer executes in the next decoding step, which the default 30\% update ratio allows.

\begin{figure}[t]
    \centering
    \includegraphics[width=0.95\linewidth]{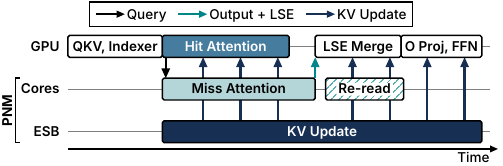}
    \vspace{-5pt}
    \caption{Execution timeline of \sys.}
    \label{fig:execution_timeline}
    % \vspace{-5pt}
\end{figure}

% \begingroup
% \makeatletter
% \appto\fs@ruled{\appto\@fs@post{\vspace{-5pt}}}
% \makeatother

\begin{algorithm}[t]
\caption{Runtime cache-factor selection}
\label{alg:adaptive_cache}
\begin{algorithmic}[1]
\Require Active requests $A$, pending queue $Q$, calibrated table $T$, HBM capacity $M$
\Statex $\textproc{Mem}(B, C)$: HBM required by $B$ requests at cache factor $C$, including weights and indexer metadata
\State $B \gets \min(|A| + |Q|,\; B^{*})$ \Comment{candidate batch size}
\While{$\textproc{Mem}(B, C_{\min}) > M$ \textbf{and} $B > |A|$}
    \State $B \gets B - 1$ \Comment{defer the newest candidate}
\EndWhile
\State $C \gets T[B]$ \Comment{calibrated cache factor for $B$}
\While{$\textproc{Mem}(B, C) > M$}
    \State lower $C$ by one step \Comment{highest factor that fits}
\EndWhile
\State resize the GPU caches of $A$ to $C$
\State admit $B - |A|$ requests from $Q$ in FIFO order
\end{algorithmic}
\end{algorithm}

% \endgroup

\subsection{KV Read Reuse with Early Streaming Buffer}
\label{sec:system_streaming}

GPU cache update candidates are a subset of the same KV entries that the PNM already reads to compute miss attention.
Our update path therefore captures candidate data from the existing attention read stream instead of fetching the same KV entries through a separate memory access.

\noindent\textbf{Capture and stream.}
As the PNM reads miss KV from local memory, the update path identifies update candidates and copies their data into an on-chip Early Streaming Buffer (ESB).
Because miss attention reads keys when computing attention scores ($QK^\top$) and reads values later when aggregating them, keys and values are captured separately as KV fragments.
Each fragment is streamed to the GPU as soon as it is captured, without waiting for its key or value counterpart or for miss attention to finish.
Attention-result transfers take priority over cache-update traffic on the \mbox{GPU-facing link.}

\noindent\textbf{Rate decoupling.}
The ESB separates the high-rate PNM memory-read path from the lower-bandwidth GPU-facing path.
This allows PNM attention to continue consuming KV at its local-memory rate while cache-update traffic drains independently toward the GPU.
The ESB therefore serves as temporary rate-decoupling storage rather than as another level of the KV cache.

\noindent\textbf{Finite-buffer handling.}
A candidate KV fragment can miss capture for two reasons: the ESB is full, or the read is served from the PNM's own cache and never appears on the DRAM read path.
\sys therefore does not rely on capture alone.
After miss attention completes, it compares the registered update candidates against the fragments actually captured and fetches the missing ones with explicit DRAM reads, so a finite ESB never stalls miss attention while every update candidate still reaches the GPU (Figure~\ref{fig:execution_timeline}).

% Figure~\ref{fig:execution_timeline} highlights how PNM attention, GPU-cache updates, and attention-result transfers overlap in this execution path.
% The concrete PNM-side structures implementing candidate identification, buffering, and completion tracking are described in \S\ref{sec:system_pnm_arch}.

\subsection{Adaptive Cache Factor}
\label{sec:system_adaptive}

Following Principle~3 (\S\ref{sec:principle_adaptive}), \sys lowers the cache factor under high load to admit pending requests into the batch, and raises it again as the load decreases.
\sys first performs offline calibration to build a table $T$ that maps each batch size to a high-throughput cache factor, together with the throughput-optimal batch size $B^{*}$.
At every decoding-step boundary, the host scheduler runs Algorithm~\ref{alg:adaptive_cache}: it admits as many pending requests as fit in HBM at the minimum cache factor without evicting active ones (lines~1--4), and then sets the calibrated factor for the resulting batch, lowering it until it fits (lines~5--8).
Increasing the factor reserves GPU cache capacity that subsequent cache updates fill, and decreasing it evicts entries once pending updates to the affected requests complete.
The algorithm itself is lightweight, involving only table lookups and simple HBM-footprint checks per decoding step.

\begin{figure}[t]
    \centering
    \includegraphics[width=0.95\linewidth]{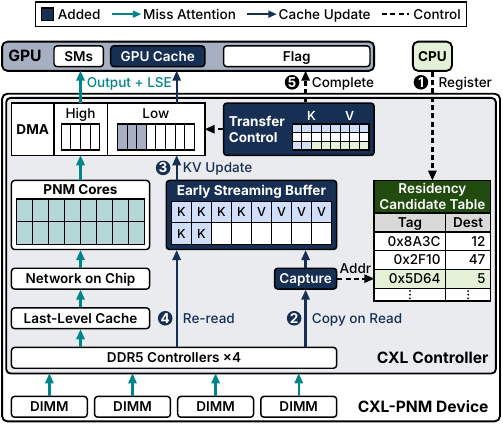}
    \vspace{-5pt}
    \caption{PNM microarchitecture of \sys.}
    \label{fig:pnm_microarchitecture}
    \vspace{-18pt}
\end{figure}
\begin{figure*}[t]
    \centering
    \includegraphics[width=0.95\textwidth]{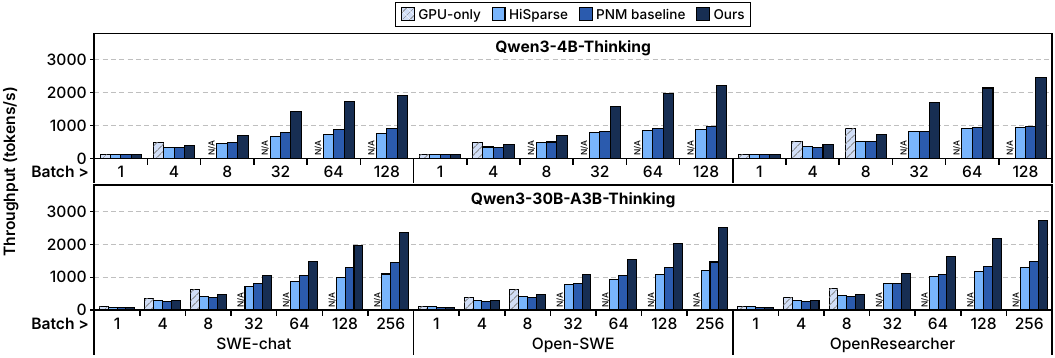}
    \caption{Decode throughput comparison across different batch sizes.}
    \label{fig:decode_throughput}
    \vspace{-10pt}
\end{figure*}
\subsection{PNM Microarchitecture}
\label{sec:system_pnm_arch}

The read-reusing update path requires only limited extensions to the existing CXL-PNM organization.
We retain the existing processing cores, coherent interconnect, cache hierarchy, memory controllers, DRAM channels, CXL controller, and DMA infrastructure, and augment the PNM with three logical functions: candidate identification and capture, buffering and transmission, and re-read and completion tracking.
Figure~\ref{fig:pnm_microarchitecture} shows these additions and their placement relative to the existing PNM memory path.

\noindent\textbf{Candidate identification and capture.}
Before miss attention begins, the host registers update candidates in a \emph{Residency Candidate Table} (RCT); each entry holds the address tag of a candidate and its destination in the GPU cache.
The capture logic taps the read-return path of the memory controllers, so it reuses the data already returning to the cores and issues no extra DRAM read.
For each returned cache line, it looks up the address in the RCT; on a match, it writes a copy into the ESB and marks its fragment as captured in a per-update bitmap.
The capture logic makes no policy decision and only checks membership in the host-registered set.
It also never back-pressures the attention data path: when the ESB lacks space, only the copy is dropped and the original read proceeds.

\noindent\textbf{Buffering and transmission.}
The ESB, an 8\,MiB on-chip SRAM, implements the rate decoupling of \S\ref{sec:system_streaming}; we choose this capacity because it captures nearly all update data under the default 30\% update ratio (\S\ref{sec:evaluation_overhead}).
Transfer-control logic tracks ready KV fragments and forwards them through the existing DMA path as GPU-facing transfer resources become available.
ESB space can be released once the DMA engine has read the fragment.
Transfer control issues cache-update transfers on a lower-priority DMA channel than attention results.
Our system assumes platform support for PCIe P2P DMA between the CXL-PNM device and GPU memory. 

\noindent\textbf{Re-read and completion tracking.}
After miss attention for the current layer completes, the PNM identifies the uncaptured fragments from the bitmap, regardless of whether the capture logic ever observed them.
It then issues explicit re-reads that bypass the PNM cache hierarchy and fetch these fragments from DRAM directly into reserved ESB space; otherwise, a fragment still residing in a cache, such as the LLC, would be served from there and never reach the capture point.
Our evaluation charges each re-read a full DRAM read without any cache benefit.
Once all fragments of an update have been transferred, the PNM writes a per-layer completion flag to GPU memory, which the runtime checks before the next execution of that layer (\S\ref{sec:system_selective_update}).
This mechanism preserves correct cache state even when the ESB is smaller than the peak update demand.
\section{Methodology}
\label{sec:methodology}

\noindent\textbf{Hardware configuration.}
We evaluate Qwen3-4B-Thinking and Qwen3-30B-A3B-Thinking~\cite{qwen3} on one and two NVIDIA H100 NVL GPUs, respectively, with an Intel Xeon 6520P host.
Following prior work~\cite{nelssa}, we model four CXL-PNM devices, each with four DDR5-6400 channels (200~GB/s) and 512~GB of memory, forming a 2-TB offloading pool.
HiSparse~\cite{hisparse} uses a host-memory pool of the same capacity, and both use the same PCIe Gen5$\times$16 link to the GPU.

\noindent\textbf{Workloads and sparse traces.}
We use three long-context agentic workloads: SWE-chat, Open-SWE, and OpenResearcher~\cite{swechat,openswe,openresearcher}.
To characterize selection locality and bounded cache updates, we train indexers following SeerAttention-R~\cite{seerattention_r}, SAS~\cite{sas}, and DSA~\cite{deepseek_v32} and collect their layer- and head-level selection traces on all three workloads.
For system evaluation, we replay SeerAttention-R traces with a Top-2K budget along each request's decoding progression; all systems receive the same requests and traces.

\noindent\textbf{Baselines.}
\emph{GPU-only} keeps the full KV cache in HBM and runs all sparse attention on the GPU (N/A when HBM capacity is exceeded).
\emph{HiSparse}~\cite{hisparse}, a state-of-the-art KV offloading system for sparse attention, keeps a GPU cache backed by host memory and fetches misses to the GPU before attention.
\emph{PNM baseline} keeps only a small set of steady tokens on the GPU and runs most of sparse attention on the PNM~\cite{scalablepnm}.
\emph{\sys} uses a 30\% recency-prioritized update ratio and an 8~MiB ESB per PNM device (\S\ref{sec:system_overview}).

\noindent\textbf{System modeling.}
We model the serving system with LLMServingSim~\cite{llmservingsim}, covering request queueing and admission, GPU cache allocation under HBM limits, GPU--PNM execution, link contention, cache-update traffic, finite ESB capacity, and re-reads.
GPU kernel latencies (indexing, attention, LSE merging, projections, and FFNs) are profiled on H100 NVL, PNM latencies are derived from M2NDP~\cite{m2ndp} configured to these devices, and CPU-side overheads are measured on the host.

\niparagraph{Hardware cost modeling.}
We model the ESB, RCT, and transfer control (including the capture bitmap) with CACTI 7.0~\cite{cacti} at 22~nm, scale area and per-access energy to 7~nm with DeepScaleTool~\cite{deepscaletool}, and estimate dynamic power from per-workload access counts.  
% workload-dependent access counts.

\section{Experimental Results}
\label{sec:evaluation}

\subsection{Offline Decode Throughput Evaluation}
\label{sec:evaluation_decode_throughput}

\niparagraph{Experimental setup.}
We sweep the batch size up to 128 for Qwen3-4B-Thinking and 256 for Qwen3-30B-A3B-Thinking on all three workloads, sampling each request's context length from the workload's distribution and fixing its output length to the workload's median.
To isolate the benefit of GPU--PNM execution from adaptive cache allocation, HiSparse and \sys use a fixed cache factor of 1.0.
Decode throughput is the number of generated tokens divided by the time to complete the batch.

\niparagraph{Results.}
Figure~\ref{fig:decode_throughput} shows the decode throughput of the four systems.
GPU-only is competitive at small batch sizes but soon exceeds HBM capacity, because it keeps the full KV cache of every request in HBM.
HiSparse and the PNM baseline sustain larger batches, yet their throughput saturates as the GPU sits idle: HiSparse waits for every miss to arrive over the PCIe link before attention, and the PNM baseline waits for the lower-bandwidth PNM to compute most of sparse attention.
\sys instead keeps a high hit ratio and computes the remaining misses on the PNM, which removes miss transfers from the critical path, eliminating the PCIe bottleneck that limits HiSparse, while leaving far less work to the PNM than the PNM baseline.
At the largest batch sizes, \sys achieves 2.6$\times$ and 2.1$\times$ higher throughput than HiSparse on Qwen3-4B-Thinking and Qwen3-30B-A3B-Thinking, respectively, and 2.3$\times$ and 1.7$\times$ higher than the PNM baseline, averaged over the three workloads.
% Completion Latency (latency knee, 5 req/s ~ 20 req/s)
% P50,  P99 TPOT  (plot over 5 req/s ~ 20 req/s)
\begin{figure}[t]
    \centering
    \includegraphics[width=0.9\linewidth]{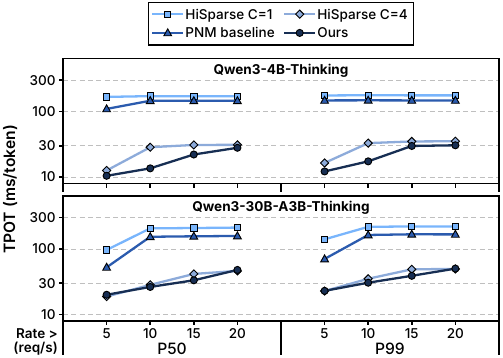}
    \caption{TPOT comparison under different arrival rates.}
    \label{fig:online_tpot}
    \vspace{-10pt}
\end{figure}
\subsection{Online Serving Evaluation}
\label{sec:evaluation_online}

\begin{figure}[t]
    \centering
    \includegraphics[width=0.9\linewidth]{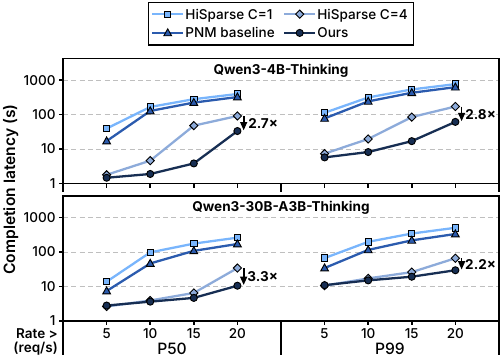}
    \caption{Completion latency under different arrival rates.}
    \label{fig:online_completion_latency}
    \vspace{-10pt}
\end{figure}

\niparagraph{Experimental setup.}
We evaluate online decode serving on Qwen3-4B-Thinking and Qwen3-30B-A3B-Thinking using SWE-chat.
Each request samples its context length and output length from the SWE-chat trace distribution, so the online workload reflects the length variation of real agentic sessions.
Requests arrive at the decode instance following a Poisson process at 5, 10, 15, and 20~requests/s, with their prompt KV already available in the backing memory.
For each arrival rate, requests are injected for 200~s, and the system runs until all of them complete; requests that cannot be admitted immediately wait in the queue.
We compare \sys against the PNM baseline and HiSparse with fixed cache factors of 1.0, 2.0, 3.0, and 4.0.
For clarity, the figures show HiSparse with cache factors of 1.0 and 4.0, the latter being the best-performing baseline.
All systems receive identical requests, arrival times, sparse-selection traces, context lengths, and output lengths.

\begin{figure}[t]
    \centering
    \includegraphics[width=0.95\linewidth]{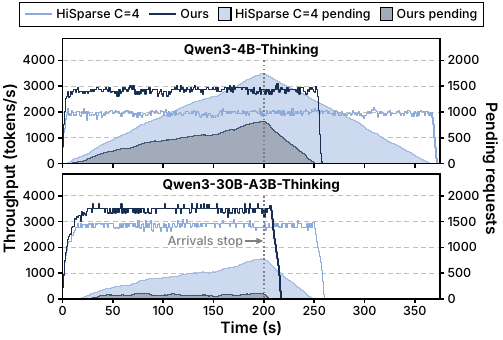}
    \vspace{-5pt}
    \caption{Online serving over time at 20~requests/s.}
    \label{fig:online_decode_throughput}
    \vspace{-10pt}
\end{figure}

\niparagraph{Metrics.}
We report three metrics to separate execution efficiency from queueing effects.
Time per output token (TPOT) is the average interval between consecutive output tokens of a request, measured from its first to its last generated token.
Completion latency is measured from arrival at the decode instance to the final output token, and therefore includes the time a request waits for admission.
To show how queueing builds up and drains, we also trace the number of pending requests over time, together with the number of tokens generated per second.

\niparagraph{TPOT.}
Figure~\ref{fig:online_tpot} shows that \sys reduces P50 TPOT by up to 16.0$\times$ and 7.8$\times$ over HiSparse with a cache factor of 1.0 on Qwen3-4B-Thinking and Qwen3-30B-A3B-Thinking, respectively, and by up to 10.8$\times$ and 5.8$\times$ over the PNM baseline.
Against HiSparse with a cache factor of 4.0, which allocates a large GPU cache to each request, \sys reduces P50 TPOT by up to 2.1$\times$ at low to moderate arrival rates.
At higher arrival rates, \sys lowers its cache factor to admit more requests, so its TPOT approaches that of HiSparse at 20~requests/s while it decodes 1.2--1.3$\times$ more requests concurrently.

\niparagraph{Completion latency.}
Figure~\ref{fig:online_completion_latency} shows that the gap between the systems widens once admission waiting time is counted.
At low arrival rates, all systems experience little queueing.
As the arrival rate increases, the fixed-cache HiSparse configurations accumulate waiting requests: a large cache factor limits how many requests fit in HBM, whereas a small one increases the misses that must be fetched over PCIe before attention.
\sys avoids this trade-off because the PNM processes the misses with internal memory bandwidth far exceeding that of the PCIe link, so \sys can admit more requests with a smaller cache factor under high load.
At 20~requests/s, \sys reduces P50 and P99 completion latency by 2.7$\times$ and 2.8$\times$ on Qwen3-4B-Thinking, and by 3.3$\times$ and 2.2$\times$ on Qwen3-30B-A3B-Thinking, over the best-performing baseline.

\niparagraph{Online serving over time.}
Figure~\ref{fig:online_decode_throughput} explains these gains by tracing pending requests over time at 20~requests/s, comparing \sys against HiSparse with a cache factor of 4.0, the best-performing baseline.
While requests arrive, \sys computes misses on the PNM instead of fetching them over PCIe before attention, so it completes requests faster and keeps fewer of them waiting.
The number of pending requests thus peaks 2.1$\times$ and 6.6$\times$ lower with \sys on Qwen3-4B-Thinking and Qwen3-30B-A3B-Thinking, respectively, and its queue drains 115~s and 43~s earlier after arrivals stop.
HiSparse, in contrast, waits for miss transfers at every step, so requests accumulate for as long as they keep arriving.

\subsection{Sensitivity Analysis}
\label{sec:evaluation_sensitivity}
% Static GPU-PNM V.S. Ours 
\begin{figure}[t]
    \centering
    \includegraphics[width=0.9\linewidth]{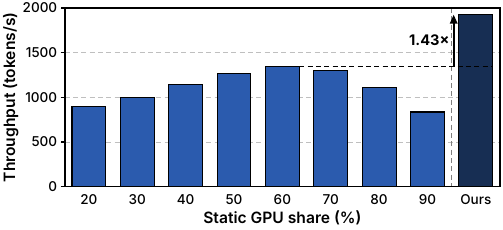}
    \vspace{-6pt}
    \caption{\sys versus static GPU--PNM partitions.}
    \vspace{-6pt}
    \label{fig:vs_static_gpu_pnm}
\end{figure}
\begin{figure}[t]
    \centering
    \includegraphics[width=0.9\linewidth]{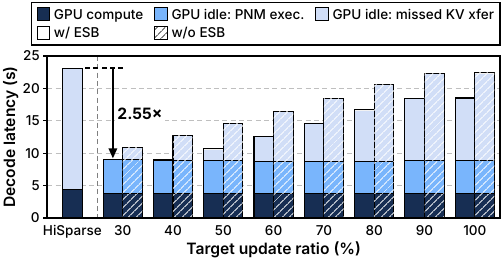}
    \vspace{-5pt}
    \caption{Decode latency breakdown across update ratios.}
    \vspace{-15pt}
    \label{fig:latency_breakdown}
\end{figure}

We report the sensitivity analysis on Qwen3-4B-Thinking with SWE-chat under the offline setup of \S\ref{sec:evaluation_decode_throughput} at its largest batch size; Qwen3-30B-A3B-Thinking shows similar trends.

\niparagraph{Static GPU--PNM partition.}
Figure~\ref{fig:vs_static_gpu_pnm} compares \sys against static partitions that assign a fixed share of each selection, from 20\% to 90\%, to the GPU.
Raising the GPU share first improves throughput, because less sparse attention is left to the lower-bandwidth PNM.
Beyond a 60\% share, however, throughput declines: keeping a larger share of each selection on the GPU requires more KV entries to cross the GPU-facing link, and the GPU idles while waiting for these transfers.
Even the best static partition remains 1.43$\times$ slower than \sys, so tuning a fixed split is not enough.
\sys instead lets selection locality set the GPU share and computes the remaining misses on the PNM, so no KV transfer lies on the critical path of the current step; bounded cache updates limit the link traffic  needed to sustain this locality, and the ESB avoids redundant DRAM reads.

\begin{figure}[t]
    \centering
    \includegraphics[width=0.93\linewidth]{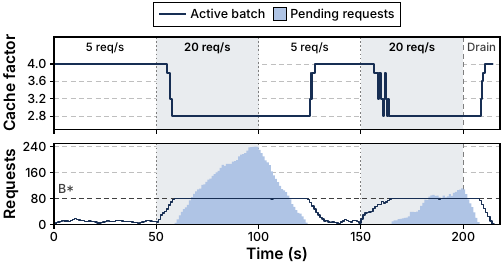}
    \vspace{-7pt}
    \caption{Adaptive cache factor under alternating load.}
    \label{fig:acf_timeline}
    \vspace{-10pt}
\end{figure}

\begin{figure}[t]
    \centering
    \includegraphics[width=0.9\linewidth]{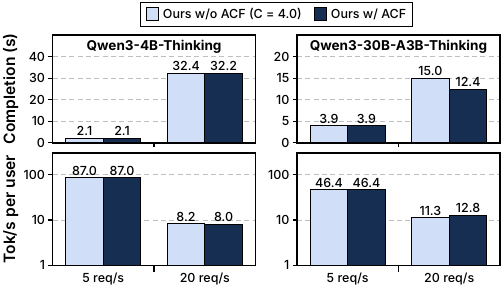}
    \vspace{-5pt}
    \caption{Completion latency and per-user throughput of \sys with and without the adaptive cache factor (ACF).}
    \label{fig:acf_vs_fixed}
    \vspace{-13pt}
\end{figure}

% Latency breakdown (HiSparse, 100% update ~ 30% update)
\niparagraph{Latency breakdown.}
Figure~\ref{fig:latency_breakdown} breaks decode latency into GPU computation, GPU idle time waiting for miss attention, and GPU idle time waiting for KV transfers.
At a 30\% update ratio, \sys reduces decode latency by 2.55$\times$ over HiSparse, because computing misses on the PNM removes their transfer from the current step.
Raising the update ratio adds link traffic that the GPU must wait for, and removing the ESB adds further delay from re-reads.
These results support the 30\% update ratio (\S\ref{sec:principle_selective_update}) and the ESB (\S\ref{sec:principle_data_reuse}).

\subsection{Ablation Study}
\label{sec:evaluation_ablation}

\niparagraph{Adaptive cache factor behavior.}
Figure~\ref{fig:acf_timeline} traces the cache factor as the load alternates between 5 and 20~requests/s every 50~s on Qwen3-4B-Thinking with SWE-chat; Qwen3-30B-A3B-Thinking behaves similarly.
Under low load, \sys keeps the cache factor at 4.0, serving most of each selection from the GPU cache.
At 20~requests/s, it lowers the cache factor to 2.8 to admit more requests until the active batch reaches $B^*$, the throughput-optimal batch size (\S\ref{sec:system_adaptive}); growing the batch further would cut the cache factor so far that too much attention falls on the PNM, lowering throughput.
Excess requests wait in the queue, and \sys holds the lower cache factor until they drain, even after the load drops to 5~requests/s at 100~s; only then does it restore the factor to 4.0.

\niparagraph{Difference from fixed cache factor.}
Figure~\ref{fig:acf_vs_fixed} compares \sys with a variant that fixes the cache factor at 4.0.
At 5~requests/s, the adaptive cache factor selects 4.0 as well, so both perform identically.
At 20~requests/s, it lowers the cache factor to admit a larger batch.
On Qwen3-4B-Thinking, the two configurations remain comparable.
On Qwen3-30B-A3B-Thinking, whose larger weights make batching more beneficial, the larger batch raises per-user throughput by 1.13$\times$ and reduces average completion latency by 1.21$\times$.
\subsection{Hardware Overhead}
\label{sec:evaluation_overhead}

\begin{figure}[t]
    \centering
    \includegraphics[width=0.9\linewidth]{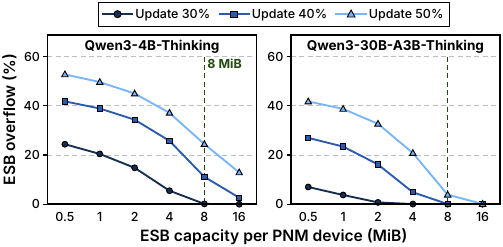}
    \vspace{-5pt}
    \caption{ESB overflow versus its capacity.}
    \label{fig:buffer_size_sensitivity}
    \vspace{-10pt}
\end{figure}

\begin{table}[t]
    \centering
    \caption{Hardware overhead per PNM device.}
    \label{tab:pnm_hardware_overhead}
    \small
    \setlength{\tabcolsep}{4pt}
    \begin{tabular}{lrr}
        \toprule
        \textbf{Component} & \textbf{Area (mm$^2$)} & \textbf{Power (mW)} \\
        \midrule
        Early Streaming Buffer (ESB)     & 1.6981 & 111.12 \\
        Residency Candidate Table (RCT)  & 0.1007 & 59.58  \\
        Transfer control                & 0.2107 & 15.14  \\
        \midrule
        \textbf{Total}                   & \textbf{2.0095} & \textbf{185.84} \\
        \bottomrule
    \end{tabular}
\end{table}

\niparagraph{ESB capacity.}
The ESB capacity determines how much update data is captured during miss attention and how much must be re-read from DRAM.
Figure~\ref{fig:buffer_size_sensitivity} shows the share of update data that overflows the ESB for update ratios of 30\%, 40\%, and 50\%, averaged over the three workloads at a cache factor of 1.0 and the largest batch size of each model (\S\ref{sec:evaluation_decode_throughput}), where update traffic is highest.
With the default 30\% update ratio, an 8~MiB ESB captures nearly all update data on both models, whereas at 40\% and 50\% the same capacity lets 11.1\% and 24.4\% overflow on Qwen3-4B-Thinking.
The ESB capacity and the update ratio are thus sized together: because Principle~1 (\S\ref{sec:principle_selective_update}) bounds the update volume, 8~MiB per device suffices, and we use this capacity throughout the evaluation.

\niparagraph{Area and power.}
Table~\ref{tab:pnm_hardware_overhead} lists the area and average dynamic power of the structures that \sys adds to each PNM device.
The additions occupy 2.01~mm$^2$ and consume 186~mW per device, or 8.04~mm$^2$ and 0.74~W across the four devices, with the ESB accounting for most of the area.
Marvell Structera A is manufactured in a 5-nm process~\cite{marvell_structera_sk_hynix_2026}, so our 7-nm estimates are conservative.

\section{Related Work}
\label{sec:related_work}

\subsection{Sparse Attention and KV Cache Offloading}
\label{sec:related_work_gpu_system}

Sparse attention restricts each query to a small subset of KV entries.
Unlike KV eviction~\cite{h2o, streamingllm, snapkv} and quantization~\cite{kivi, kvquant}, which discard or approximate entries, it must keep the full KV cache accessible.
The selection is either applied only at inference time~\cite{quest, infinigen, retroinfer, arkvale, retrievalattention, pqcache} or incorporated into model training~\cite{nsa, seerattention_r, deepseek_v32, sas, infllmv2, moba, glm5}; \sys consumes only the resulting selections and is orthogonal to the selection method, including trained indexers.

KV offloading moves the KV cache to host memory during decoding~\cite{flexgen, ess} or to host memory and storage for reuse across requests~\cite{mooncake, cachedattention, strata}.
For sparse attention, HiSparse~\cite{hisparse} and related systems~\cite{arkvale,infllm,spin,pulseinfer,nosa} keep recently selected entries in a GPU cache and fetch every miss from host memory before attention.
ECHO~\cite{echo} and ShadowKV~\cite{shadowkv} hide or reduce these transfers, but every miss must still reach the GPU before attention.
In these systems, a GPU cache update is a transfer required for the current step; in \sys, the PNM computes the current misses, so updates serve only future locality and their volume becomes a design choice.

\subsection{Near-Memory Processing for LLM Inference}
\label{sec:related_work_ndp_system}

PIM and PNM architectures accelerate memory-bound decoding operators such as attention and state updates~\cite{attacc,neupims,pimba,pimphony,l3,cxlpnm}, and M2NDP~\cite{m2ndp} enables general-purpose near-data processing in CXL memory expanders.
Recent GPU--PNM systems apply near-memory processing to long-context sparse attention.
KARAT~\cite{karat} executes both the indexer and sparse attention on the PNM while the GPU runs the remaining operators, LongSight~\cite{longsight} keeps recent KV entries on the GPU and offloads Top-$k$ scoring of older entries to a CXL memory device, and NELSSA~\cite{nelssa} places each request on the GPU or the PNM according to its length.
These systems share a PNM-centric execution, in which the PNM processes most of the sparse attention and the GPU takes a share fixed by operator type, token position, or request length.

ScalablePNM~\cite{scalablepnm} also places part of the KV cache on the GPU, but differs from \sys in three ways.
First, its hybrid mode keeps a small set of steady tokens on the GPU at a static ratio, leaving most of the attention to the PNM; in \sys, the GPU computes most of the attention, following the cache hits of each step.
Second, tokens moved to its GPU must arrive before the current step's attention, whereas \sys computes current misses on the PNM, so its updates serve only future locality; \sys identifies the saturation point of this locality and updates just enough to sustain a high hit ratio.
Third, sustaining a GPU-major share requires continuous updates over the narrow link, so \sys captures update data from the PNM's attention reads with the ESB and overlaps their transfer with PNM attention.

\section{Conclusion}
\label{sec:conclusion}

Sparse attention leaves the capacity demand of long-context decoding unchanged.
PNM holds the full KV cache and computes where it resides, but its lower bandwidth than GPU HBM leaves the GPU idle under PNM-centric execution.
Exploiting the selection locality of sparse attention, \sys instead makes GPU--PNM execution GPU-centric.
The GPU computes most of the attention from its cache, while the PNM computes the misses.
Because misses need not reach the GPU before attention, \sys turns GPU cache updates into investments in future locality, bounds them where the hit ratio saturates, captures them from the PNM's reads with the Early Streaming Buffer, and adapts the GPU cache size to the serving load.
\sys improves decode throughput by 2.3$\times$ over HiSparse and 2.0$\times$ over the PNM baseline on average.
These results show that PNM is most effective not as the primary attention engine but as a computing tier of the GPU memory hierarchy.
%%%%%%% -- PAPER CONTENT ENDS -- %%%%%%%%

\begin{acks}
This work was supported by SK hynix Inc.\ for the development of the heterogeneous GPU--PNM LLM serving system, and by the Institute of Information \& Communications Technology Planning \& Evaluation (IITP) grant funded by the Korea government (MSIT) (No. RS-2026-25548502, Development of Adaptive System Software Technology for AI Ecosystems Enabling Continuous Performance Optimization of AI Semiconductors) for extending the LLMServingSim simulator to model this system.
\end{acks}

\bibliographystyle{ACM-Reference-Format}
\bibliography{refs}


\begin{thebibliography}{71}

%%% except for \shownote{} and \showURL{}.  The latter does
%%% not use final punctuation, in order to avoid confusing it with
%%% the Web address.

\ifx \showCODEN    \undefined \def \showCODEN     #1{\unskip}     \fi
\ifx \showISBNx    \undefined \def \showISBNx     #1{\unskip}     \fi
\ifx \showISBNxiii \undefined \def \showISBNxiii  #1{\unskip}     \fi
\ifx \showISSN     \undefined \def \showISSN      #1{\unskip}     \fi
\ifx \showLCCN     \undefined \def \showLCCN      #1{\unskip}     \fi
\ifx \shownote     \undefined \def \shownote      #1{#1}          \fi
\ifx \showarticletitle \undefined \def \showarticletitle #1{#1}   \fi
\ifx \showURL      \undefined \def \showURL       {\relax}        \fi
% invisible to TeX
\providecommand\bibfield[2]{#2}
\providecommand\bibinfo[2]{#2}
\providecommand\natexlab[1]{#1}
\providecommand\showeprint[2][]{arXiv:#2}
\makeatletter
\@ifundefined{NAT@parse@date}{}{\let\NAT@parse@date@orig\NAT@parse@date}
\@ifundefined{NAT@parse@date}{}{\def\NAT@parse@date#1#2#3#4#5#6@@{\NAT@parse@date@orig#1#2#3#4#5#6@@\def\NAT@tempyear{0000}\def\NAT@tempexlab{{?}}\ifx\NAT@year\NAT@tempyear\ifx\NAT@exlab\NAT@tempexlab\def\NAT@date{[n.\,d.]}\else\edef\NAT@date{[n.\,d.]\NAT@exlab}\fi\fi}}
\makeatother

\bibitem[Abhyankar et~al\mbox{.}(2024)]%
        {infercept}
\bibfield{author}{\bibinfo{person}{Reyna Abhyankar}, \bibinfo{person}{Zijian He}, \bibinfo{person}{Vikranth Srivatsa}, \bibinfo{person}{Hao Zhang}, {and} \bibinfo{person}{Yiying Zhang}.} \bibinfo{year}{2024}\natexlab{}.
\newblock \showarticletitle{{I}nfer{C}ept: Efficient Intercept Support for Augmented Large Language Model Inference}. In \bibinfo{booktitle}{\emph{Proceedings of the 41st International Conference on Machine Learning}} \emph{(\bibinfo{series}{Proceedings of Machine Learning Research}, Vol.~\bibinfo{volume}{235})}. \bibinfo{publisher}{PMLR}, \bibinfo{pages}{81--95}.
\newblock


\bibitem[Ahmad et~al\mbox{.}(2026)]%
        {openswe}
\bibfield{author}{\bibinfo{person}{Wasi Ahmad}, \bibinfo{person}{Nikolai Ludwig}, \bibinfo{person}{Somshubra Majumdar}, {and} \bibinfo{person}{Boris Ginsburg}.} \bibinfo{year}{2026}\natexlab{}.
\newblock \showarticletitle{{Open-SWE-Traces}: Advancing Dual-Mode Multilingual Distillation for Software Engineering Agents}. In \bibinfo{booktitle}{\emph{Findings of the Association for Computational Linguistics: EMNLP 2026}}. \bibinfo{pages}{2668--2679}.
\newblock


\bibitem[Balasubramonian et~al\mbox{.}(2017)]%
        {cacti}
\bibfield{author}{\bibinfo{person}{Rajeev Balasubramonian}, \bibinfo{person}{Andrew~B Kahng}, \bibinfo{person}{Naveen Muralimanohar}, \bibinfo{person}{Ali Shafiee}, {and} \bibinfo{person}{Vaishnav Srinivas}.} \bibinfo{year}{2017}\natexlab{}.
\newblock \showarticletitle{{CACTI} 7: New Tools for Interconnect Exploration in Innovative Off-Chip Memories}.
\newblock \bibinfo{journal}{\emph{ACM Transactions on Architecture and Code Optimization (TACO)}} \bibinfo{volume}{14}, \bibinfo{number}{2} (\bibinfo{year}{2017}), \bibinfo{pages}{1--25}.
\newblock


\bibitem[Baumann et~al\mbox{.}(2026)]%
        {swechat}
\bibfield{author}{\bibinfo{person}{Joachim Baumann}, \bibinfo{person}{Vishakh Padmakumar}, \bibinfo{person}{Xiang Li}, \bibinfo{person}{John Yang}, \bibinfo{person}{Diyi Yang}, {and} \bibinfo{person}{Sanmi Koyejo}.} \bibinfo{year}{2026}\natexlab{}.
\newblock \showarticletitle{{SWE-chat}: Coding Agent Interactions From Real Users in the Wild}. In \bibinfo{booktitle}{\emph{Third Conference on Language Modeling}}.
\newblock


\bibitem[Chatterji et~al\mbox{.}(2025)]%
        {chatgpt}
\bibfield{author}{\bibinfo{person}{Aaron Chatterji}, \bibinfo{person}{Thomas Cunningham}, \bibinfo{person}{David~J Deming}, \bibinfo{person}{Zoe Hitzig}, \bibinfo{person}{Christopher Ong}, \bibinfo{person}{Carl~Yan Shan}, {and} \bibinfo{person}{Kevin Wadman}.} \bibinfo{year}{2025}\natexlab{}.
\newblock \bibinfo{booktitle}{\emph{How People Use {ChatGPT}}}.
\newblock \bibinfo{type}{{T}echnical {R}eport}. \bibinfo{institution}{National Bureau of Economic Research}.
\newblock


\bibitem[Chen et~al\mbox{.}(2024)]%
        {arkvale}
\bibfield{author}{\bibinfo{person}{Renze Chen}, \bibinfo{person}{Zhuofeng Wang}, \bibinfo{person}{Beiquan Cao}, \bibinfo{person}{Tong Wu}, \bibinfo{person}{Size Zheng}, \bibinfo{person}{Xiuhong Li}, \bibinfo{person}{Xuechao Wei}, \bibinfo{person}{Shengen Yan}, \bibinfo{person}{Meng Li}, {and} \bibinfo{person}{Yun Liang}.} \bibinfo{year}{2024}\natexlab{}.
\newblock \showarticletitle{{ArkVale}: Efficient Generative {LLM} Inference with Recallable Key-Value Eviction}.
\newblock \bibinfo{journal}{\emph{Advances in Neural Information Processing Systems}}  \bibinfo{volume}{37} (\bibinfo{year}{2024}), \bibinfo{pages}{113134--113155}.
\newblock


\bibitem[Chen et~al\mbox{.}(2025)]%
        {ess}
\bibfield{author}{\bibinfo{person}{Xinhang Chen}, \bibinfo{person}{Chao Zhang}, \bibinfo{person}{Jiahuan He}, \bibinfo{person}{Wei Liu}, \bibinfo{person}{Jianming Zhang}, \bibinfo{person}{Wenlong Zhou}, \bibinfo{person}{Xiao Li}, \bibinfo{person}{Pai Zeng}, \bibinfo{person}{Shiyong Li}, \bibinfo{person}{Yuanpan Qian}, \bibinfo{person}{Dong Li}, {and} \bibinfo{person}{Zhaogeng Li}.} \bibinfo{year}{2025}\natexlab{}.
\newblock \showarticletitle{{ESS}: An Offload-Centric Latent-Cache Management Architecture for {DeepSeek-V3.2-Exp}}.
\newblock \bibinfo{journal}{\emph{arXiv preprint arXiv:2512.10576}} (\bibinfo{year}{2025}).
\newblock


\bibitem[Chen et~al\mbox{.}(2026)]%
        {retroinfer}
\bibfield{author}{\bibinfo{person}{Yaoqi Chen}, \bibinfo{person}{Jinkai Zhang}, \bibinfo{person}{Baotong Lu}, \bibinfo{person}{Qianxi Zhang}, \bibinfo{person}{Chengruidong Zhang}, \bibinfo{person}{Jing Liu}, \bibinfo{person}{Jingjia Luo}, \bibinfo{person}{Di Liu}, \bibinfo{person}{Huiqiang Jiang}, \bibinfo{person}{Qi Chen}, \bibinfo{person}{Bailu Ding}, \bibinfo{person}{Xiao Yan}, \bibinfo{person}{Jiawei Jiang}, \bibinfo{person}{Chen Chen}, \bibinfo{person}{Mingxing Zhang}, \bibinfo{person}{Cheng Li}, \bibinfo{person}{Yuqing Yang}, \bibinfo{person}{Fan Yang}, {and} \bibinfo{person}{Mao Yang}.} \bibinfo{year}{2026}\natexlab{}.
\newblock \showarticletitle{{RetroInfer}: A Vector Storage Engine for Scalable Long-Context {LLM} Inference}.
\newblock \bibinfo{journal}{\emph{Proceedings of the VLDB Endowment}} \bibinfo{volume}{19}, \bibinfo{number}{5} (\bibinfo{year}{2026}), \bibinfo{pages}{1016--1031}.
\newblock


\bibitem[Cho et~al\mbox{.}(2026)]%
        {llmservingsim}
\bibfield{author}{\bibinfo{person}{Jaehong Cho}, \bibinfo{person}{Hyunmin Choi}, \bibinfo{person}{Guseul Heo}, {and} \bibinfo{person}{Jongse Park}.} \bibinfo{year}{2026}\natexlab{}.
\newblock \showarticletitle{{LLMServingSim 2.0: A Unified Simulator for Heterogeneous and Disaggregated LLM Serving Infrastructure}}. In \bibinfo{booktitle}{\emph{2026 IEEE International Symposium on Performance Analysis of Systems and Software (ISPASS)}}. IEEE, \bibinfo{pages}{1--14}.
\newblock


\bibitem[Choi et~al\mbox{.}(2026)]%
        {nelssa}
\bibfield{author}{\bibinfo{person}{Sookyung Choi}, \bibinfo{person}{Seungyong Lee}, \bibinfo{person}{Kangkyu Park}, \bibinfo{person}{Yunseo Chun}, \bibinfo{person}{Junseok Lee}, \bibinfo{person}{Hyeongseok Gwak}, \bibinfo{person}{Myunghyun Rhee}, \bibinfo{person}{Euiseok Kim}, \bibinfo{person}{Donguk Moon}, \bibinfo{person}{Kwangsik Shin}, \bibinfo{person}{Guseul Heo}, \bibinfo{person}{Youngpyo Joo}, \bibinfo{person}{Hoshik Kim}, {and} \bibinfo{person}{Jongse Park}.} \bibinfo{year}{2026}\natexlab{}.
\newblock \showarticletitle{{NELSSA}: A {GPU-PNM} Heterogeneous System for Mixed-Length {LLM} Serving via Length-Based Request Placement}. In \bibinfo{booktitle}{\emph{Proceedings of the 59th IEEE/ACM International Symposium on Microarchitecture (MICRO)}}.
\newblock


\bibitem[{DeepSeek-AI} et~al\mbox{.}(2025)]%
        {deepseek_v32}
\bibfield{author}{\bibinfo{person}{{DeepSeek-AI}}, \bibinfo{person}{Aixin Liu}, \bibinfo{person}{Aoxue Mei}, \bibinfo{person}{Bangcai Lin}, \bibinfo{person}{Bing Xue}, \bibinfo{person}{Bingxuan Wang}, \bibinfo{person}{Bingzheng Xu}, \bibinfo{person}{Bochao Wu}, \bibinfo{person}{Bowei Zhang}, \bibinfo{person}{Chaofan Lin}, \bibinfo{person}{Chen Dong}, \bibinfo{person}{Chengda Lu}, \bibinfo{person}{Chenggang Zhao}, \bibinfo{person}{Chengqi Deng}, \bibinfo{person}{Chenhao Xu}, \bibinfo{person}{Chong Ruan}, \bibinfo{person}{Damai Dai}, \bibinfo{person}{Daya Guo}, \bibinfo{person}{Dejian Yang}, \bibinfo{person}{Deli Chen}, \bibinfo{person}{Erhang Li}, \bibinfo{person}{Fangqi Zhou}, \bibinfo{person}{Fangyun Lin}, \bibinfo{person}{Fucong Dai}, \bibinfo{person}{Guangbo Hao}, \bibinfo{person}{Guanting Chen}, \bibinfo{person}{Guowei Li}, \bibinfo{person}{H. Zhang}, \bibinfo{person}{Hanwei Xu}, \bibinfo{person}{Hao Li}, \bibinfo{person}{Haofen Liang}, \bibinfo{person}{Haoran Wei}, \bibinfo{person}{Haowei Zhang},
  \bibinfo{person}{Haowen Luo}, \bibinfo{person}{Haozhe Ji}, \bibinfo{person}{Honghui Ding}, \bibinfo{person}{Hongxuan Tang}, \bibinfo{person}{Huanqi Cao}, \bibinfo{person}{Huazuo Gao}, \bibinfo{person}{Hui Qu}, \bibinfo{person}{Hui Zeng}, \bibinfo{person}{Jialiang Huang}, \bibinfo{person}{Jiashi Li}, \bibinfo{person}{Jiaxin Xu}, \bibinfo{person}{Jiewen Hu}, \bibinfo{person}{Jingchang Chen}, \bibinfo{person}{Jingting Xiang}, \bibinfo{person}{Jingyang Yuan}, \bibinfo{person}{Jingyuan Cheng}, \bibinfo{person}{Jinhua Zhu}, \bibinfo{person}{Jun Ran}, \bibinfo{person}{Junguang Jiang}, \bibinfo{person}{Junjie Qiu}, \bibinfo{person}{Junlong Li}, \bibinfo{person}{Junxiao Song}, \bibinfo{person}{Kai Dong}, \bibinfo{person}{Kaige Gao}, \bibinfo{person}{Kang Guan}, \bibinfo{person}{Kexin Huang}, \bibinfo{person}{Kexing Zhou}, \bibinfo{person}{Kezhao Huang}, \bibinfo{person}{Kuai Yu}, \bibinfo{person}{Lean Wang}, \bibinfo{person}{Lecong Zhang}, \bibinfo{person}{Lei Wang}, \bibinfo{person}{Liang Zhao},
  \bibinfo{person}{Liangsheng Yin}, \bibinfo{person}{Lihua Guo}, \bibinfo{person}{Lingxiao Luo}, \bibinfo{person}{Linwang Ma}, \bibinfo{person}{Litong Wang}, \bibinfo{person}{Liyue Zhang}, \bibinfo{person}{M.~S. Di}, \bibinfo{person}{M.~Y Xu}, \bibinfo{person}{Mingchuan Zhang}, \bibinfo{person}{Minghua Zhang}, \bibinfo{person}{Minghui Tang}, \bibinfo{person}{Mingxu Zhou}, \bibinfo{person}{Panpan Huang}, \bibinfo{person}{Peixin Cong}, \bibinfo{person}{Peiyi Wang}, \bibinfo{person}{Qiancheng Wang}, \bibinfo{person}{Qihao Zhu}, \bibinfo{person}{Qingyang Li}, \bibinfo{person}{Qinyu Chen}, \bibinfo{person}{Qiushi Du}, \bibinfo{person}{Ruiling Xu}, \bibinfo{person}{Ruiqi Ge}, \bibinfo{person}{Ruisong Zhang}, \bibinfo{person}{Ruizhe Pan}, \bibinfo{person}{Runji Wang}, \bibinfo{person}{Runqiu Yin}, \bibinfo{person}{Runxin Xu}, \bibinfo{person}{Ruomeng Shen}, \bibinfo{person}{Ruoyu Zhang}, \bibinfo{person}{S.~H. Liu}, \bibinfo{person}{Shanghao Lu}, \bibinfo{person}{Shangyan Zhou}, \bibinfo{person}{Shanhuang Chen},
  \bibinfo{person}{Shaofei Cai}, \bibinfo{person}{Shaoyuan Chen}, \bibinfo{person}{Shengding Hu}, \bibinfo{person}{Shengyu Liu}, \bibinfo{person}{Shiqiang Hu}, \bibinfo{person}{Shirong Ma}, \bibinfo{person}{Shiyu Wang}, \bibinfo{person}{Shuiping Yu}, \bibinfo{person}{Shunfeng Zhou}, \bibinfo{person}{Shuting Pan}, \bibinfo{person}{Songyang Zhou}, \bibinfo{person}{Tao Ni}, \bibinfo{person}{Tao Yun}, \bibinfo{person}{Tian Pei}, \bibinfo{person}{Tian Ye}, \bibinfo{person}{Tianyuan Yue}, \bibinfo{person}{Wangding Zeng}, \bibinfo{person}{Wen Liu}, \bibinfo{person}{Wenfeng Liang}, \bibinfo{person}{Wenjie Pang}, \bibinfo{person}{Wenjing Luo}, \bibinfo{person}{Wenjun Gao}, \bibinfo{person}{Wentao Zhang}, \bibinfo{person}{Xi Gao}, \bibinfo{person}{Xiangwen Wang}, \bibinfo{person}{Xiao Bi}, \bibinfo{person}{Xiaodong Liu}, \bibinfo{person}{Xiaohan Wang}, \bibinfo{person}{Xiaokang Chen}, \bibinfo{person}{Xiaokang Zhang}, \bibinfo{person}{Xiaotao Nie}, \bibinfo{person}{Xin Cheng}, \bibinfo{person}{Xin Liu},
  \bibinfo{person}{Xin Xie}, \bibinfo{person}{Xingchao Liu}, \bibinfo{person}{Xingkai Yu}, \bibinfo{person}{Xingyou Li}, \bibinfo{person}{Xinyu Yang}, \bibinfo{person}{Xinyuan Li}, \bibinfo{person}{Xu Chen}, \bibinfo{person}{Xuecheng Su}, \bibinfo{person}{Xuehai Pan}, \bibinfo{person}{Xuheng Lin}, \bibinfo{person}{Xuwei Fu}, \bibinfo{person}{Y.~Q. Wang}, \bibinfo{person}{Yang Zhang}, \bibinfo{person}{Yanhong Xu}, \bibinfo{person}{Yanru Ma}, \bibinfo{person}{Yao Li}, \bibinfo{person}{Yao Li}, \bibinfo{person}{Yao Zhao}, \bibinfo{person}{Yaofeng Sun}, \bibinfo{person}{Yaohui Wang}, \bibinfo{person}{Yi Qian}, \bibinfo{person}{Yi Yu}, \bibinfo{person}{Yichao Zhang}, \bibinfo{person}{Yifan Ding}, \bibinfo{person}{Yifan Shi}, \bibinfo{person}{Yiliang Xiong}, \bibinfo{person}{Ying He}, \bibinfo{person}{Ying Zhou}, \bibinfo{person}{Yinmin Zhong}, \bibinfo{person}{Yishi Piao}, \bibinfo{person}{Yisong Wang}, \bibinfo{person}{Yixiao Chen}, \bibinfo{person}{Yixuan Tan}, \bibinfo{person}{Yixuan Wei},
  \bibinfo{person}{Yiyang Ma}, \bibinfo{person}{Yiyuan Liu}, \bibinfo{person}{Yonglun Yang}, \bibinfo{person}{Yongqiang Guo}, \bibinfo{person}{Yongtong Wu}, \bibinfo{person}{Yu Wu}, \bibinfo{person}{Yuan Cheng}, \bibinfo{person}{Yuan Ou}, \bibinfo{person}{Yuanfan Xu}, \bibinfo{person}{Yuduan Wang}, \bibinfo{person}{Yue Gong}, \bibinfo{person}{Yuhan Wu}, \bibinfo{person}{Yuheng Zou}, \bibinfo{person}{Yukun Li}, \bibinfo{person}{Yunfan Xiong}, \bibinfo{person}{Yuxiang Luo}, \bibinfo{person}{Yuxiang You}, \bibinfo{person}{Yuxuan Liu}, \bibinfo{person}{Yuyang Zhou}, \bibinfo{person}{Z.~F. Wu}, \bibinfo{person}{Z.~Z. Ren}, \bibinfo{person}{Zehua Zhao}, \bibinfo{person}{Zehui Ren}, \bibinfo{person}{Zhangli Sha}, \bibinfo{person}{Zhe Fu}, \bibinfo{person}{Zhean Xu}, \bibinfo{person}{Zhenda Xie}, \bibinfo{person}{Zhengyan Zhang}, \bibinfo{person}{Zhewen Hao}, \bibinfo{person}{Zhibin Gou}, \bibinfo{person}{Zhicheng Ma}, \bibinfo{person}{Zhigang Yan}, \bibinfo{person}{Zhihong Shao}, \bibinfo{person}{Zhixian Huang},
  \bibinfo{person}{Zhiyu Wu}, \bibinfo{person}{Zhuoshu Li}, \bibinfo{person}{Zhuping Zhang}, \bibinfo{person}{Zian Xu}, \bibinfo{person}{Zihao Wang}, \bibinfo{person}{Zihui Gu}, \bibinfo{person}{Zijia Zhu}, \bibinfo{person}{Zilin Li}, \bibinfo{person}{Zipeng Zhang}, \bibinfo{person}{Ziwei Xie}, \bibinfo{person}{Ziyi Gao}, \bibinfo{person}{Zizheng Pan}, \bibinfo{person}{Zongqing Yao}, \bibinfo{person}{Bei Feng}, \bibinfo{person}{Hui Li}, \bibinfo{person}{J.~L. Cai}, \bibinfo{person}{Jiaqi Ni}, \bibinfo{person}{Lei Xu}, \bibinfo{person}{Meng Li}, \bibinfo{person}{Ning Tian}, \bibinfo{person}{R.~J. Chen}, \bibinfo{person}{R.~L. Jin}, \bibinfo{person}{S.~S. Li}, \bibinfo{person}{Shuang Zhou}, \bibinfo{person}{Tianyu Sun}, \bibinfo{person}{X.~Q. Li}, \bibinfo{person}{Xiangyue Jin}, \bibinfo{person}{Xiaojin Shen}, \bibinfo{person}{Xiaosha Chen}, \bibinfo{person}{Xinnan Song}, \bibinfo{person}{Xinyi Zhou}, \bibinfo{person}{Y.~X. Zhu}, \bibinfo{person}{Yanping Huang}, \bibinfo{person}{Yaohui Li}, \bibinfo{person}{Yi
  Zheng}, \bibinfo{person}{Yuchen Zhu}, \bibinfo{person}{Yunxian Ma}, \bibinfo{person}{Zhen Huang}, \bibinfo{person}{Zhipeng Xu}, \bibinfo{person}{Zhongyu Zhang}, \bibinfo{person}{Dongjie Ji}, \bibinfo{person}{Jian Liang}, \bibinfo{person}{Jianzhong Guo}, \bibinfo{person}{Jin Chen}, \bibinfo{person}{Leyi Xia}, \bibinfo{person}{Miaojun Wang}, \bibinfo{person}{Mingming Li}, \bibinfo{person}{Peng Zhang}, \bibinfo{person}{Ruyi Chen}, \bibinfo{person}{Shangmian Sun}, \bibinfo{person}{Shaoqing Wu}, \bibinfo{person}{Shengfeng Ye}, \bibinfo{person}{T. Wang}, \bibinfo{person}{W.~L. Xiao}, \bibinfo{person}{Wei An}, \bibinfo{person}{Xianzu Wang}, \bibinfo{person}{Xiaowen Sun}, \bibinfo{person}{Xiaoxiang Wang}, \bibinfo{person}{Ying Tang}, \bibinfo{person}{Yukun Zha}, \bibinfo{person}{Zekai Zhang}, \bibinfo{person}{Zhe Ju}, \bibinfo{person}{Zhen Zhang}, {and} \bibinfo{person}{Zihua Qu}.} \bibinfo{year}{2025}\natexlab{}.
\newblock \showarticletitle{{DeepSeek-V3.2}: Pushing the Frontier of Open Large Language Models}.
\newblock \bibinfo{journal}{\emph{arXiv preprint arXiv:2512.02556}} (\bibinfo{year}{2025}).
\newblock


\bibitem[{DeepSeek-AI} et~al\mbox{.}(2026)]%
        {deepseek_v4}
\bibfield{author}{\bibinfo{person}{{DeepSeek-AI}}, \bibinfo{person}{Anyi Xu}, \bibinfo{person}{Bangcai Lin}, \bibinfo{person}{Bing Xue}, \bibinfo{person}{Bingxuan Wang}, \bibinfo{person}{Bingzheng Xu}, \bibinfo{person}{Bochao Wu}, \bibinfo{person}{Bowei Zhang}, \bibinfo{person}{Chaofan Lin}, \bibinfo{person}{Chen Dong}, \bibinfo{person}{Chenchen Ling}, \bibinfo{person}{Chengda Lu}, \bibinfo{person}{Chenggang Zhao}, \bibinfo{person}{Chengqi Deng}, \bibinfo{person}{Chengyu Hou}, \bibinfo{person}{Chenhao Xu}, \bibinfo{person}{Chenze Shao}, \bibinfo{person}{Chong Ruan}, \bibinfo{person}{Conner Sun}, \bibinfo{person}{Damai Dai}, \bibinfo{person}{Daya Guo}, \bibinfo{person}{Dejian Yang}, \bibinfo{person}{Deli Chen}, \bibinfo{person}{Donghao Li}, \bibinfo{person}{Dongjie Ji}, \bibinfo{person}{Erhang Li}, \bibinfo{person}{Fang Wei}, \bibinfo{person}{Fangyun Lin}, \bibinfo{person}{Fangzhou Yuan}, \bibinfo{person}{Feiyu Xia}, \bibinfo{person}{Fucong Dai}, \bibinfo{person}{Guangbo Hao}, \bibinfo{person}{Guanting Chen},
  \bibinfo{person}{Guoai Cao}, \bibinfo{person}{Guolai Meng}, \bibinfo{person}{Guowei Li}, \bibinfo{person}{Han Yu}, \bibinfo{person}{Han Zhang}, \bibinfo{person}{Hanwei Xu}, \bibinfo{person}{Hao Li}, \bibinfo{person}{Haofen Liang}, \bibinfo{person}{Haoling Zhang}, \bibinfo{person}{Haoming Luo}, \bibinfo{person}{Haoran Wei}, \bibinfo{person}{Haotian Yuan}, \bibinfo{person}{Haowei Zhang}, \bibinfo{person}{Haowen Luo}, \bibinfo{person}{Haoyu Chen}, \bibinfo{person}{Haozhe Ji}, \bibinfo{person}{Hengqing Zhang}, \bibinfo{person}{Honghui Ding}, \bibinfo{person}{Hongxuan Tang}, \bibinfo{person}{Huanqi Cao}, \bibinfo{person}{Huazuo Gao}, \bibinfo{person}{Hui Qu}, \bibinfo{person}{Hui Zeng}, \bibinfo{person}{J Yang}, \bibinfo{person}{JQ Zhu}, \bibinfo{person}{Jia Luo}, \bibinfo{person}{Jia Song}, \bibinfo{person}{Jia Yu}, \bibinfo{person}{Jialiang Huang}, \bibinfo{person}{Jialu Cai}, \bibinfo{person}{Jian Liang}, \bibinfo{person}{Jiangting Zhou}, \bibinfo{person}{Jiasheng Ye}, \bibinfo{person}{Jiashi Li},
  \bibinfo{person}{Jiaxin Xu}, \bibinfo{person}{Jiewen Hu}, \bibinfo{person}{Jieyu Yang}, \bibinfo{person}{Jin Chen}, \bibinfo{person}{Jin Yan}, \bibinfo{person}{Jingchang Chen}, \bibinfo{person}{Jingli Zhou}, \bibinfo{person}{Jingting Xiang}, \bibinfo{person}{Jingyang Yuan}, \bibinfo{person}{Jingyuan Cheng}, \bibinfo{person}{Jingzi Zhou}, \bibinfo{person}{Jinhua Zhu}, \bibinfo{person}{Jiping Yu}, \bibinfo{person}{Joseph Sun}, \bibinfo{person}{Jun Ran}, \bibinfo{person}{Junguang Jiang}, \bibinfo{person}{Junjie Qiu}, \bibinfo{person}{Junlong Li}, \bibinfo{person}{Junmin Zheng}, \bibinfo{person}{Junxiao Song}, \bibinfo{person}{Kai Dong}, \bibinfo{person}{Kaige Gao}, \bibinfo{person}{Kang Guan}, \bibinfo{person}{Kexing Zhou}, \bibinfo{person}{Kezhao Huang}, \bibinfo{person}{Kuai Yu}, \bibinfo{person}{Lean Wang}, \bibinfo{person}{Lecong Zhang}, \bibinfo{person}{Lei Wang}, \bibinfo{person}{Leyi Xia}, \bibinfo{person}{Li Zhang}, \bibinfo{person}{Liang Zhao}, \bibinfo{person}{Lihua Guo}, \bibinfo{person}{Lingxiao
  Luo}, \bibinfo{person}{Linwang Ma}, \bibinfo{person}{Linyan Zhu}, \bibinfo{person}{Litong Wang}, \bibinfo{person}{Liyu Cai}, \bibinfo{person}{Liyue Zhang}, \bibinfo{person}{Longhao Chen}, \bibinfo{person}{MS Di}, \bibinfo{person}{MY Xu}, \bibinfo{person}{Max Mei}, \bibinfo{person}{Miaojun Wang}, \bibinfo{person}{Mingchuan Zhang}, \bibinfo{person}{Minghua Zhang}, \bibinfo{person}{Minghui Tang}, \bibinfo{person}{Mingming Li}, \bibinfo{person}{Mingxu Zhou}, \bibinfo{person}{Minmin Han}, \bibinfo{person}{Ning Wang}, \bibinfo{person}{Panpan Huang}, \bibinfo{person}{Panpan Wang}, \bibinfo{person}{Peixin Cong}, \bibinfo{person}{Peiyi Wang}, \bibinfo{person}{Peng Zhang}, \bibinfo{person}{Qiancheng Wang}, \bibinfo{person}{Qihao Zhu}, \bibinfo{person}{Qingyang Li}, \bibinfo{person}{Qinyu Chen}, \bibinfo{person}{Qiushi Du}, \bibinfo{person}{Qiwei Jiang}, \bibinfo{person}{Rui Tian}, \bibinfo{person}{Ruifan Xu}, \bibinfo{person}{Ruijie Lu}, \bibinfo{person}{Ruiling Xu}, \bibinfo{person}{Ruiqi Ge},
  \bibinfo{person}{Ruisong Zhang}, \bibinfo{person}{Ruizhe Pan}, \bibinfo{person}{Runji Wang}, \bibinfo{person}{Runqian Chen}, \bibinfo{person}{Runqiu Yin}, \bibinfo{person}{Runxin Xu}, \bibinfo{person}{Ruomeng Shen}, \bibinfo{person}{Ruoyu Zhang}, \bibinfo{person}{Ruyi Chen}, \bibinfo{person}{SH Liu}, \bibinfo{person}{Shanghao Lu}, \bibinfo{person}{Shangmian Sun}, \bibinfo{person}{Shangyan Zhou}, \bibinfo{person}{Shanhuang Chen}, \bibinfo{person}{Shaofei Cai}, \bibinfo{person}{Shaoheng Nie}, \bibinfo{person}{Shaoqing Wu}, \bibinfo{person}{Shaoyuan Chen}, \bibinfo{person}{Shengding Hu}, \bibinfo{person}{Shengyu Liu}, \bibinfo{person}{Shiqiang Hu}, \bibinfo{person}{Shirong Ma}, \bibinfo{person}{Shiyu Wang}, \bibinfo{person}{Shuiping Yu}, \bibinfo{person}{Shunfeng Zhou}, \bibinfo{person}{Shuting Pan}, \bibinfo{person}{Shuying Yu}, \bibinfo{person}{Songyang Zhou}, \bibinfo{person}{Tao Ni}, \bibinfo{person}{Tao Yun}, \bibinfo{person}{Tian Jin}, \bibinfo{person}{Tian Pei}, \bibinfo{person}{Tian Ye},
  \bibinfo{person}{Tianle Lin}, \bibinfo{person}{Tianran Ji}, \bibinfo{person}{Tianyi Cui}, \bibinfo{person}{Tianyuan Yue}, \bibinfo{person}{Tingting Yu}, \bibinfo{person}{Tun Wang}, \bibinfo{person}{W Zhang}, \bibinfo{person}{WL Xiao}, \bibinfo{person}{Wangding Zeng}, \bibinfo{person}{Wei An}, \bibinfo{person}{Weilin Zhao}, \bibinfo{person}{Wen Liu}, \bibinfo{person}{Wenfeng Liang}, \bibinfo{person}{Wenjie Pang}, \bibinfo{person}{Wenjing Luo}, \bibinfo{person}{Wenjing Yao}, \bibinfo{person}{Wenjun Gao}, \bibinfo{person}{Wenkai Yang}, \bibinfo{person}{Wenlve Huang}, \bibinfo{person}{Wenqing Hou}, \bibinfo{person}{Wentao Zhang}, \bibinfo{person}{Wenting Ma}, \bibinfo{person}{Xi Gao}, \bibinfo{person}{Xiang He}, \bibinfo{person}{Xiangwen Wang}, \bibinfo{person}{Xianzu Wang}, \bibinfo{person}{Xiao Bi}, \bibinfo{person}{Xiaodong Liu}, \bibinfo{person}{Xiaohan Wang}, \bibinfo{person}{Xiaokang Chen}, \bibinfo{person}{Xiaokang Zhang}, \bibinfo{person}{Xiaotao Nie}, \bibinfo{person}{Xiaowen Sun},
  \bibinfo{person}{Xiaoxiang Wang}, \bibinfo{person}{Xin Cheng}, \bibinfo{person}{Xin Liu}, \bibinfo{person}{Xin Xie}, \bibinfo{person}{Xingchao Liu}, \bibinfo{person}{Xingchen Liu}, \bibinfo{person}{Xingkai Yu}, \bibinfo{person}{Xingyou Li}, \bibinfo{person}{Xinyu Yang}, \bibinfo{person}{Xinyu Zhang}, \bibinfo{person}{Xu Chen}, \bibinfo{person}{Xuanyu Wang}, \bibinfo{person}{Xuecheng Su}, \bibinfo{person}{Xueyin Chen}, \bibinfo{person}{Xuheng Lin}, \bibinfo{person}{Xuwei Fu}, \bibinfo{person}{YC Yan}, \bibinfo{person}{YQ Wang}, \bibinfo{person}{YW Ma}, \bibinfo{person}{Yanfeng Luo}, \bibinfo{person}{Yang Zhang}, \bibinfo{person}{Yanhong Xu}, \bibinfo{person}{Yanru Ma}, \bibinfo{person}{Yanwen Huang}, \bibinfo{person}{Yao Li}, \bibinfo{person}{Yao Li}, \bibinfo{person}{Yao Xu}, \bibinfo{person}{Yao Zhao}, \bibinfo{person}{Yaofeng Sun}, \bibinfo{person}{Yaohui Wang}, \bibinfo{person}{Yi Qian}, \bibinfo{person}{Yi Shao}, \bibinfo{person}{Yi Yu}, \bibinfo{person}{Yichao Zhang}, \bibinfo{person}{Yifan Ding},
  \bibinfo{person}{Yifan Shi}, \bibinfo{person}{Yijia Wu}, \bibinfo{person}{Yiliang Xiong}, \bibinfo{person}{Yiling Ma}, \bibinfo{person}{Ying He}, \bibinfo{person}{Ying Tang}, \bibinfo{person}{Ying Zhou}, \bibinfo{person}{Yingjia Luo}, \bibinfo{person}{Yinmin Zhong}, \bibinfo{person}{Yishi Piao}, \bibinfo{person}{Yisong Wang}, \bibinfo{person}{Yixiang Zhang}, \bibinfo{person}{Yixiao Chen}, \bibinfo{person}{Yixuan Tan}, \bibinfo{person}{Yixuan Wei}, \bibinfo{person}{Yiyang Ma}, \bibinfo{person}{Yiyuan Liu}, \bibinfo{person}{Yonglun Yang}, \bibinfo{person}{Yongqiang Guo}, \bibinfo{person}{Yongtong Wu}, \bibinfo{person}{Yu Wu}, \bibinfo{person}{YuKun Li}, \bibinfo{person}{Yuan Cheng}, \bibinfo{person}{Yuan Ou}, \bibinfo{person}{Yuanfan Xu}, \bibinfo{person}{Yuanhao Li}, \bibinfo{person}{Yuduan Wang}, \bibinfo{person}{Yuehan Yang}, \bibinfo{person}{Yuer Xu}, \bibinfo{person}{Yuhan Wu}, \bibinfo{person}{Yuhao Meng}, \bibinfo{person}{Yuheng Zou}, \bibinfo{person}{Yukun Zha}, \bibinfo{person}{Yunfan Xiong},
  \bibinfo{person}{Yupeng Chen}, \bibinfo{person}{Yuping Lin}, \bibinfo{person}{Yuqian Cao}, \bibinfo{person}{Yuqian Wang}, \bibinfo{person}{Yushun Zhang}, \bibinfo{person}{Yuting Yan}, \bibinfo{person}{Yutong Lin}, \bibinfo{person}{Yuxian Gu}, \bibinfo{person}{Yuxiang Luo}, \bibinfo{person}{Yuxiang You}, \bibinfo{person}{Yuxuan Liu}, \bibinfo{person}{Yuxuan Zhou}, \bibinfo{person}{Yuyang Zhou}, \bibinfo{person}{Yuzhen Huang}, \bibinfo{person}{ZF Wu}, \bibinfo{person}{Zehao Wang}, \bibinfo{person}{Zehua Zhao}, \bibinfo{person}{Zehui Ren}, \bibinfo{person}{Zekai Zhang}, \bibinfo{person}{Zhangli Sha}, \bibinfo{person}{Zhe Fu}, \bibinfo{person}{Zhe Ju}, \bibinfo{person}{Zhean Xu}, \bibinfo{person}{Zhenda Xie}, \bibinfo{person}{Zhengyan Zhang}, \bibinfo{person}{Zheren Gao}, \bibinfo{person}{Zhewen Hao}, \bibinfo{person}{Zhibin Gou}, \bibinfo{person}{Zhicheng Ma}, \bibinfo{person}{Zhigang Yan}, \bibinfo{person}{Zhihong Shao}, \bibinfo{person}{Zhixian Huang}, \bibinfo{person}{Zhixuan Chen}, \bibinfo{person}{Zhiyu
  Wu}, \bibinfo{person}{Zhizhou Ren}, \bibinfo{person}{Zhongyu Wu}, \bibinfo{person}{Zhuoshu Li}, \bibinfo{person}{Zhuping Zhang}, \bibinfo{person}{Zian Xu}, \bibinfo{person}{Zihao Wang}, \bibinfo{person}{Zihua Qu}, \bibinfo{person}{Zihui Gu}, \bibinfo{person}{Zijia Zhu}, \bibinfo{person}{Zilin Li}, \bibinfo{person}{Zipeng Zhang}, \bibinfo{person}{Ziwei Xie}, \bibinfo{person}{Ziyi Gao}, \bibinfo{person}{Ziyi Wan}, \bibinfo{person}{Zizheng Pan}, {and} \bibinfo{person}{Zongqing Yao}.} \bibinfo{year}{2026}\natexlab{}.
\newblock \showarticletitle{{DeepSeek-V4}: Towards Highly Efficient Million-Token Context Intelligence}.
\newblock \bibinfo{journal}{\emph{arXiv preprint arXiv:2606.19348}} (\bibinfo{year}{2026}).
\newblock


\bibitem[Gao et~al\mbox{.}(2024)]%
        {cachedattention}
\bibfield{author}{\bibinfo{person}{Bin Gao}, \bibinfo{person}{Zhuomin He}, \bibinfo{person}{Puru Sharma}, \bibinfo{person}{Qingxuan Kang}, \bibinfo{person}{Djordje Jevdjic}, \bibinfo{person}{Junbo Deng}, \bibinfo{person}{Xingkun Yang}, \bibinfo{person}{Zhou Yu}, {and} \bibinfo{person}{Pengfei Zuo}.} \bibinfo{year}{2024}\natexlab{}.
\newblock \showarticletitle{Cost-Efficient Large Language Model Serving for Multi-Turn Conversations with {CachedAttention}}. In \bibinfo{booktitle}{\emph{2024 USENIX annual technical conference (USENIX ATC 24)}}. \bibinfo{pages}{111--126}.
\newblock


\bibitem[Gao et~al\mbox{.}(2026)]%
        {seerattention_r}
\bibfield{author}{\bibinfo{person}{Yizhao Gao}, \bibinfo{person}{Shuming Guo}, \bibinfo{person}{Shijie Cao}, \bibinfo{person}{Yuqing Xia}, \bibinfo{person}{Yu Cheng}, \bibinfo{person}{Lei Wang}, \bibinfo{person}{Lingxiao Ma}, \bibinfo{person}{Yutao Sun}, \bibinfo{person}{Tianzhu Ye}, \bibinfo{person}{Li Dong}, \bibinfo{person}{Hayden Kwok-Hay So}, \bibinfo{person}{Yu Hua}, \bibinfo{person}{Ting Cao}, \bibinfo{person}{Fan Yang}, {and} \bibinfo{person}{Mao Yang}.} \bibinfo{year}{2026}\natexlab{}.
\newblock \showarticletitle{Sparse Attention Adaptation for Long Reasoning}. In \bibinfo{booktitle}{\emph{International Conference on Learning Representations}}, Vol.~\bibinfo{volume}{2026}. \bibinfo{pages}{111697--111714}.
\newblock


\bibitem[{GLM-5-Team} et~al\mbox{.}(2026)]%
        {glm5}
\bibfield{author}{\bibinfo{person}{{GLM-5-Team}}, \bibinfo{person}{Aohan Zeng}, \bibinfo{person}{Xin Lv}, \bibinfo{person}{Zhenyu Hou}, \bibinfo{person}{Zhengxiao Du}, \bibinfo{person}{Qinkai Zheng}, \bibinfo{person}{Bin Chen}, \bibinfo{person}{Da Yin}, \bibinfo{person}{Chendi Ge}, \bibinfo{person}{Chenghua Huang}, \bibinfo{person}{Chengxing Xie}, \bibinfo{person}{Chenzheng Zhu}, \bibinfo{person}{Congfeng Yin}, \bibinfo{person}{Cunxiang Wang}, \bibinfo{person}{Gengzheng Pan}, \bibinfo{person}{Hao Zeng}, \bibinfo{person}{Haoke Zhang}, \bibinfo{person}{Haoran Wang}, \bibinfo{person}{Huilong Chen}, \bibinfo{person}{Jiajie Zhang}, \bibinfo{person}{Jian Jiao}, \bibinfo{person}{Jiaqi Guo}, \bibinfo{person}{Jingsen Wang}, \bibinfo{person}{Jingzhao Du}, \bibinfo{person}{Jinzhu Wu}, \bibinfo{person}{Kedong Wang}, \bibinfo{person}{Lei Li}, \bibinfo{person}{Lin Fan}, \bibinfo{person}{Lucen Zhong}, \bibinfo{person}{Mingdao Liu}, \bibinfo{person}{Mingming Zhao}, \bibinfo{person}{Pengfan Du}, \bibinfo{person}{Qian Dong},
  \bibinfo{person}{Rui Lu}, \bibinfo{person}{{Shuang-Li}}, \bibinfo{person}{Shulin Cao}, \bibinfo{person}{Song Liu}, \bibinfo{person}{Ting Jiang}, \bibinfo{person}{Xiaodong Chen}, \bibinfo{person}{Xiaohan Zhang}, \bibinfo{person}{Xuancheng Huang}, \bibinfo{person}{Xuezhen Dong}, \bibinfo{person}{Yabo Xu}, \bibinfo{person}{Yao Wei}, \bibinfo{person}{Yifan An}, \bibinfo{person}{Yilin Niu}, \bibinfo{person}{Yitong Zhu}, \bibinfo{person}{Yuanhao Wen}, \bibinfo{person}{Yukuo Cen}, \bibinfo{person}{Yushi Bai}, \bibinfo{person}{Zhongpei Qiao}, \bibinfo{person}{Zihan Wang}, \bibinfo{person}{Zikang Wang}, \bibinfo{person}{Zilin Zhu}, \bibinfo{person}{Ziqiang Liu}, \bibinfo{person}{Zixuan Li}, \bibinfo{person}{Bojie Wang}, \bibinfo{person}{Bosi Wen}, \bibinfo{person}{Can Huang}, \bibinfo{person}{Changpeng Cai}, \bibinfo{person}{Chao Yu}, \bibinfo{person}{Chen Li}, \bibinfo{person}{Chengwei Hu}, \bibinfo{person}{Chenhui Zhang}, \bibinfo{person}{Dan Zhang}, \bibinfo{person}{Daoyan Lin}, \bibinfo{person}{Dayong Yang},
  \bibinfo{person}{Di Wang}, \bibinfo{person}{Ding Ai}, \bibinfo{person}{Erle Zhu}, \bibinfo{person}{Fangzhou Yi}, \bibinfo{person}{Feiyu Chen}, \bibinfo{person}{Guohong Wen}, \bibinfo{person}{Hailong Sun}, \bibinfo{person}{Haisha Zhao}, \bibinfo{person}{Haiyi Hu}, \bibinfo{person}{Hanchen Zhang}, \bibinfo{person}{Hanrui Liu}, \bibinfo{person}{Hanyu Zhang}, \bibinfo{person}{Hao Peng}, \bibinfo{person}{Hao Tai}, \bibinfo{person}{Haobo Zhang}, \bibinfo{person}{He Liu}, \bibinfo{person}{Hongwei Wang}, \bibinfo{person}{Hongxi Yan}, \bibinfo{person}{Hongyu Ge}, \bibinfo{person}{Huan Liu}, \bibinfo{person}{Huanpeng Chu}, \bibinfo{person}{Jia'ni Zhao}, \bibinfo{person}{Jiachen Wang}, \bibinfo{person}{Jiajing Zhao}, \bibinfo{person}{Jiamin Ren}, \bibinfo{person}{Jiapeng Wang}, \bibinfo{person}{Jiaxin Zhang}, \bibinfo{person}{Jiayi Gui}, \bibinfo{person}{Jiayue Zhao}, \bibinfo{person}{Jijie Li}, \bibinfo{person}{Jing An}, \bibinfo{person}{Jing Li}, \bibinfo{person}{Jingwei Yuan}, \bibinfo{person}{Jinhua Du},
  \bibinfo{person}{Jinxin Liu}, \bibinfo{person}{Junkai Zhi}, \bibinfo{person}{Junwen Duan}, \bibinfo{person}{Kaiyue Zhou}, \bibinfo{person}{Kangjian Wei}, \bibinfo{person}{Ke Wang}, \bibinfo{person}{Keyun Luo}, \bibinfo{person}{Laiqiang Zhang}, \bibinfo{person}{Leigang Sha}, \bibinfo{person}{Liang Xu}, \bibinfo{person}{Lindong Wu}, \bibinfo{person}{Lintao Ding}, \bibinfo{person}{Lu Chen}, \bibinfo{person}{Minghao Li}, \bibinfo{person}{Nianyi Lin}, \bibinfo{person}{Pan Ta}, \bibinfo{person}{Qiang Zou}, \bibinfo{person}{Rongjun Song}, \bibinfo{person}{Ruiqi Yang}, \bibinfo{person}{Shangqing Tu}, \bibinfo{person}{Shangtong Yang}, \bibinfo{person}{Shaoxiang Wu}, \bibinfo{person}{Shengyan Zhang}, \bibinfo{person}{Shijie Li}, \bibinfo{person}{Shuang Li}, \bibinfo{person}{Shuyi Fan}, \bibinfo{person}{Wei Qin}, \bibinfo{person}{Wei Tian}, \bibinfo{person}{Weining Zhang}, \bibinfo{person}{Wenbo Yu}, \bibinfo{person}{Wenjie Liang}, \bibinfo{person}{Xiang Kuang}, \bibinfo{person}{Xiangmeng Cheng},
  \bibinfo{person}{Xiangyang Li}, \bibinfo{person}{Xiaoquan Yan}, \bibinfo{person}{Xiaowei Hu}, \bibinfo{person}{Xiaoying Ling}, \bibinfo{person}{Xing Fan}, \bibinfo{person}{Xingye Xia}, \bibinfo{person}{Xinyuan Zhang}, \bibinfo{person}{Xinze Zhang}, \bibinfo{person}{Xirui Pan}, \bibinfo{person}{Xu Zou}, \bibinfo{person}{Xunkai Zhang}, \bibinfo{person}{Yadi Liu}, \bibinfo{person}{Yandong Wu}, \bibinfo{person}{Yanfu Li}, \bibinfo{person}{Yidong Wang}, \bibinfo{person}{Yifan Zhu}, \bibinfo{person}{Yijun Tan}, \bibinfo{person}{Yilin Zhou}, \bibinfo{person}{Yiming Pan}, \bibinfo{person}{Ying Zhang}, \bibinfo{person}{Yinpei Su}, \bibinfo{person}{Yipeng Geng}, \bibinfo{person}{Yong Yan}, \bibinfo{person}{Yonglin Tan}, \bibinfo{person}{Yuean Bi}, \bibinfo{person}{Yuhan Shen}, \bibinfo{person}{Yuhao Yang}, \bibinfo{person}{Yujiang Li}, \bibinfo{person}{Yunan Liu}, \bibinfo{person}{Yunqing Wang}, \bibinfo{person}{Yuntao Li}, \bibinfo{person}{Yurong Wu}, \bibinfo{person}{Yutao Zhang}, \bibinfo{person}{Yuxi Duan},
  \bibinfo{person}{Yuxuan Zhang}, \bibinfo{person}{Zezhen Liu}, \bibinfo{person}{Zhengtao Jiang}, \bibinfo{person}{Zhenhe Yan}, \bibinfo{person}{Zheyu Zhang}, \bibinfo{person}{Zhixiang Wei}, \bibinfo{person}{Zhuo Chen}, \bibinfo{person}{Zhuoer Feng}, \bibinfo{person}{Zijun Yao}, \bibinfo{person}{Ziwei Chai}, \bibinfo{person}{Ziyuan Wang}, \bibinfo{person}{Zuzhou Zhang}, \bibinfo{person}{Bin Xu}, \bibinfo{person}{Minlie Huang}, \bibinfo{person}{Hongning Wang}, \bibinfo{person}{Juanzi Li}, \bibinfo{person}{Yuxiao Dong}, {and} \bibinfo{person}{Jie Tang}.} \bibinfo{year}{2026}\natexlab{}.
\newblock \showarticletitle{{GLM-5}: From Vibe Coding to Agentic Engineering}.
\newblock \bibinfo{journal}{\emph{arXiv preprint arXiv:2602.15763}} (\bibinfo{year}{2026}).
\newblock


\bibitem[Gottweis et~al\mbox{.}(2026)]%
        {ai_scientist}
\bibfield{author}{\bibinfo{person}{Juraj Gottweis}, \bibinfo{person}{Wei-Hung Weng}, \bibinfo{person}{Alexander Daryin}, \bibinfo{person}{Tao Tu}, \bibinfo{person}{Petar Sirkovic}, \bibinfo{person}{Artiom Myaskovsky}, \bibinfo{person}{Grzegorz Glowaty}, \bibinfo{person}{Felix Weissenberger}, \bibinfo{person}{Alessio Orlandi}, \bibinfo{person}{Dan Popovici}, \bibinfo{person}{Anil Palepu}, \bibinfo{person}{Keran Rong}, \bibinfo{person}{Ryutaro Tanno}, \bibinfo{person}{Khaled Saab}, \bibinfo{person}{Fan Zhang}, \bibinfo{person}{Jacob Blum}, \bibinfo{person}{Andrew Carroll}, \bibinfo{person}{Kavita Kulkarni}, \bibinfo{person}{Nenad Toma{\v{s}}ev}, \bibinfo{person}{Dina Zverinski}, \bibinfo{person}{Ivor Rendulic}, \bibinfo{person}{Elahe Vedadi}, \bibinfo{person}{Florian Hasler}, \bibinfo{person}{Luka Rimanic}, \bibinfo{person}{Marina Boia}, \bibinfo{person}{Ivan Budiselic}, \bibinfo{person}{Ben Feinstein}, \bibinfo{person}{Mathias Bellaiche}, \bibinfo{person}{Tom Sheffer}, \bibinfo{person}{Jan Freyberg},
  \bibinfo{person}{Jeremy Ratcliff}, \bibinfo{person}{Ottavia Bertolli}, \bibinfo{person}{Katherine Chou}, \bibinfo{person}{Avinatan Hassidim}, \bibinfo{person}{Burak Gokturk}, \bibinfo{person}{Amin Vahdat}, \bibinfo{person}{Yuan Guan}, \bibinfo{person}{Vikram Dhillon}, \bibinfo{person}{Eeshit~Dhaval Vaishnav}, \bibinfo{person}{Byron Lee}, \bibinfo{person}{Tiago R.~D. Costa}, \bibinfo{person}{Jos{\'e}~R. Penad{\'e}s}, \bibinfo{person}{Gary Peltz}, \bibinfo{person}{Yossi Matias}, \bibinfo{person}{James Manyika}, \bibinfo{person}{Demis Hassabis}, \bibinfo{person}{Yunhan Xu}, \bibinfo{person}{Pushmeet Kohli}, \bibinfo{person}{Annalisa Pawlosky}, \bibinfo{person}{Alan Karthikesalingam}, {and} \bibinfo{person}{Vivek Natarajan}.} \bibinfo{year}{2026}\natexlab{}.
\newblock \showarticletitle{Accelerating Scientific Discovery with {Co-Scientist}}.
\newblock \bibinfo{journal}{\emph{Nature}} \bibinfo{volume}{655}, \bibinfo{number}{8122} (\bibinfo{year}{2026}), \bibinfo{pages}{487--496}.
\newblock


\bibitem[Guo et~al\mbox{.}(2025)]%
        {deepseek_r1}
\bibfield{author}{\bibinfo{person}{Daya Guo}, \bibinfo{person}{Dejian Yang}, \bibinfo{person}{Haowei Zhang}, \bibinfo{person}{Junxiao Song}, \bibinfo{person}{Peiyi Wang}, \bibinfo{person}{Qihao Zhu}, \bibinfo{person}{Runxin Xu}, \bibinfo{person}{Ruoyu Zhang}, \bibinfo{person}{Shirong Ma}, \bibinfo{person}{Xiao Bi}, \bibinfo{person}{Xiaokang Zhang}, \bibinfo{person}{Xingkai Yu}, \bibinfo{person}{Yu Wu}, \bibinfo{person}{Z.~F. Wu}, \bibinfo{person}{Zhibin Gou}, \bibinfo{person}{Zhihong Shao}, \bibinfo{person}{Zhuoshu Li}, \bibinfo{person}{Ziyi Gao}, \bibinfo{person}{Aixin Liu}, \bibinfo{person}{Bing Xue}, \bibinfo{person}{Bingxuan Wang}, \bibinfo{person}{Bochao Wu}, \bibinfo{person}{Bei Feng}, \bibinfo{person}{Chengda Lu}, \bibinfo{person}{Chenggang Zhao}, \bibinfo{person}{Chengqi Deng}, \bibinfo{person}{Chong Ruan}, \bibinfo{person}{Damai Dai}, \bibinfo{person}{Deli Chen}, \bibinfo{person}{Dongjie Ji}, \bibinfo{person}{Erhang Li}, \bibinfo{person}{Fangyun Lin}, \bibinfo{person}{Fucong Dai},
  \bibinfo{person}{Fuli Luo}, \bibinfo{person}{Guangbo Hao}, \bibinfo{person}{Guanting Chen}, \bibinfo{person}{Guowei Li}, \bibinfo{person}{H. Zhang}, \bibinfo{person}{Hanwei Xu}, \bibinfo{person}{Honghui Ding}, \bibinfo{person}{Huazuo Gao}, \bibinfo{person}{Hui Qu}, \bibinfo{person}{Hui Li}, \bibinfo{person}{Jianzhong Guo}, \bibinfo{person}{Jiashi Li}, \bibinfo{person}{Jingchang Chen}, \bibinfo{person}{Jingyang Yuan}, \bibinfo{person}{Jinhao Tu}, \bibinfo{person}{Junjie Qiu}, \bibinfo{person}{Junlong Li}, \bibinfo{person}{J.~L. Cai}, \bibinfo{person}{Jiaqi Ni}, \bibinfo{person}{Jian Liang}, \bibinfo{person}{Jin Chen}, \bibinfo{person}{Kai Dong}, \bibinfo{person}{Kai Hu}, \bibinfo{person}{Kaichao You}, \bibinfo{person}{Kaige Gao}, \bibinfo{person}{Kang Guan}, \bibinfo{person}{Kexin Huang}, \bibinfo{person}{Kuai Yu}, \bibinfo{person}{Lean Wang}, \bibinfo{person}{Lecong Zhang}, \bibinfo{person}{Liang Zhao}, \bibinfo{person}{Litong Wang}, \bibinfo{person}{Liyue Zhang}, \bibinfo{person}{Lei Xu},
  \bibinfo{person}{Leyi Xia}, \bibinfo{person}{Mingchuan Zhang}, \bibinfo{person}{Minghua Zhang}, \bibinfo{person}{Minghui Tang}, \bibinfo{person}{Mingxu Zhou}, \bibinfo{person}{Meng Li}, \bibinfo{person}{Miaojun Wang}, \bibinfo{person}{Mingming Li}, \bibinfo{person}{Ning Tian}, \bibinfo{person}{Panpan Huang}, \bibinfo{person}{Peng Zhang}, \bibinfo{person}{Qiancheng Wang}, \bibinfo{person}{Qinyu Chen}, \bibinfo{person}{Qiushi Du}, \bibinfo{person}{Ruiqi Ge}, \bibinfo{person}{Ruisong Zhang}, \bibinfo{person}{Ruizhe Pan}, \bibinfo{person}{Runji Wang}, \bibinfo{person}{R.~J. Chen}, \bibinfo{person}{R.~L. Jin}, \bibinfo{person}{Ruyi Chen}, \bibinfo{person}{Shanghao Lu}, \bibinfo{person}{Shangyan Zhou}, \bibinfo{person}{Shanhuang Chen}, \bibinfo{person}{Shengfeng Ye}, \bibinfo{person}{Shiyu Wang}, \bibinfo{person}{Shuiping Yu}, \bibinfo{person}{Shunfeng Zhou}, \bibinfo{person}{Shuting Pan}, \bibinfo{person}{S.~S. Li}, \bibinfo{person}{Shuang Zhou}, \bibinfo{person}{Shaoqing Wu}, \bibinfo{person}{Tao Yun},
  \bibinfo{person}{Tian Pei}, \bibinfo{person}{Tianyu Sun}, \bibinfo{person}{T. Wang}, \bibinfo{person}{Wangding Zeng}, \bibinfo{person}{Wen Liu}, \bibinfo{person}{Wenfeng Liang}, \bibinfo{person}{Wenjun Gao}, \bibinfo{person}{Wenqin Yu}, \bibinfo{person}{Wentao Zhang}, \bibinfo{person}{W.~L. Xiao}, \bibinfo{person}{Wei An}, \bibinfo{person}{Xiaodong Liu}, \bibinfo{person}{Xiaohan Wang}, \bibinfo{person}{Xiaokang Chen}, \bibinfo{person}{Xiaotao Nie}, \bibinfo{person}{Xin Cheng}, \bibinfo{person}{Xin Liu}, \bibinfo{person}{Xin Xie}, \bibinfo{person}{Xingchao Liu}, \bibinfo{person}{Xinyu Yang}, \bibinfo{person}{Xinyuan Li}, \bibinfo{person}{Xuecheng Su}, \bibinfo{person}{Xuheng Lin}, \bibinfo{person}{X.~Q. Li}, \bibinfo{person}{Xiangyue Jin}, \bibinfo{person}{Xiaojin Shen}, \bibinfo{person}{Xiaosha Chen}, \bibinfo{person}{Xiaowen Sun}, \bibinfo{person}{Xiaoxiang Wang}, \bibinfo{person}{Xinnan Song}, \bibinfo{person}{Xinyi Zhou}, \bibinfo{person}{Xianzu Wang}, \bibinfo{person}{Xinxia Shan},
  \bibinfo{person}{Y.~K. Li}, \bibinfo{person}{Y.~Q. Wang}, \bibinfo{person}{Y.~X. Wei}, \bibinfo{person}{Yang Zhang}, \bibinfo{person}{Yanhong Xu}, \bibinfo{person}{Yao Li}, \bibinfo{person}{Yao Zhao}, \bibinfo{person}{Yaofeng Sun}, \bibinfo{person}{Yaohui Wang}, \bibinfo{person}{Yi Yu}, \bibinfo{person}{Yichao Zhang}, \bibinfo{person}{Yifan Shi}, \bibinfo{person}{Yiliang Xiong}, \bibinfo{person}{Ying He}, \bibinfo{person}{Yishi Piao}, \bibinfo{person}{Yisong Wang}, \bibinfo{person}{Yixuan Tan}, \bibinfo{person}{Yiyang Ma}, \bibinfo{person}{Yiyuan Liu}, \bibinfo{person}{Yongqiang Guo}, \bibinfo{person}{Yuan Ou}, \bibinfo{person}{Yuduan Wang}, \bibinfo{person}{Yue Gong}, \bibinfo{person}{Yuheng Zou}, \bibinfo{person}{Yujia He}, \bibinfo{person}{Yunfan Xiong}, \bibinfo{person}{Yuxiang Luo}, \bibinfo{person}{Yuxiang You}, \bibinfo{person}{Yuxuan Liu}, \bibinfo{person}{Yuyang Zhou}, \bibinfo{person}{Y.~X. Zhu}, \bibinfo{person}{Yanping Huang}, \bibinfo{person}{Yaohui Li}, \bibinfo{person}{Yi Zheng},
  \bibinfo{person}{Yuchen Zhu}, \bibinfo{person}{Yunxian Ma}, \bibinfo{person}{Ying Tang}, \bibinfo{person}{Yukun Zha}, \bibinfo{person}{Yuting Yan}, \bibinfo{person}{Z.~Z. Ren}, \bibinfo{person}{Zehui Ren}, \bibinfo{person}{Zhangli Sha}, \bibinfo{person}{Zhe Fu}, \bibinfo{person}{Zhean Xu}, \bibinfo{person}{Zhenda Xie}, \bibinfo{person}{Zhengyan Zhang}, \bibinfo{person}{Zhewen Hao}, \bibinfo{person}{Zhicheng Ma}, \bibinfo{person}{Zhigang Yan}, \bibinfo{person}{Zhiyu Wu}, \bibinfo{person}{Zihui Gu}, \bibinfo{person}{Zijia Zhu}, \bibinfo{person}{Zijun Liu}, \bibinfo{person}{Zilin Li}, \bibinfo{person}{Ziwei Xie}, \bibinfo{person}{Ziyang Song}, \bibinfo{person}{Zizheng Pan}, \bibinfo{person}{Zhen Huang}, \bibinfo{person}{Zhipeng Xu}, \bibinfo{person}{Zhongyu Zhang}, {and} \bibinfo{person}{Zhen Zhang}.} \bibinfo{year}{2025}\natexlab{}.
\newblock \showarticletitle{{DeepSeek-R1} Incentivizes Reasoning in {LLMs} through Reinforcement Learning}.
\newblock \bibinfo{journal}{\emph{Nature}} \bibinfo{volume}{645}, \bibinfo{number}{8081} (\bibinfo{year}{2025}), \bibinfo{pages}{633--638}.
\newblock


\bibitem[Ham et~al\mbox{.}(2026)]%
        {karat}
\bibfield{author}{\bibinfo{person}{Hyungkyu Ham}, \bibinfo{person}{Junhyeong Bae}, \bibinfo{person}{Seungheon Lee}, \bibinfo{person}{Myeongjae Jeon}, {and} \bibinfo{person}{Gwangsun Kim}.} \bibinfo{year}{2026}\natexlab{}.
\newblock \showarticletitle{Heterogeneous LLM Serving with General-Purpose Processing-Near-Memory for Retrieval-Based Sparse Attention}.
\newblock \bibinfo{journal}{\emph{arXiv preprint arXiv:2608.03555}} (\bibinfo{year}{2026}).
\newblock


\bibitem[Ham et~al\mbox{.}(2024)]%
        {m2ndp}
\bibfield{author}{\bibinfo{person}{Hyungkyu Ham}, \bibinfo{person}{Jeongmin Hong}, \bibinfo{person}{Geonwoo Park}, \bibinfo{person}{Yunseon Shin}, \bibinfo{person}{Okkyun Woo}, \bibinfo{person}{Wonhyuk Yang}, \bibinfo{person}{Jinhoon Bae}, \bibinfo{person}{Eunhyeok Park}, \bibinfo{person}{Hyojin Sung}, \bibinfo{person}{Euicheol Lim}, {and} \bibinfo{person}{Gwangsun Kim}.} \bibinfo{year}{2024}\natexlab{}.
\newblock \showarticletitle{Low-Overhead General-Purpose Near-Data Processing in {CXL} Memory Expanders}. In \bibinfo{booktitle}{\emph{2024 57th IEEE/ACM International Symposium on Microarchitecture (MICRO)}}. IEEE, \bibinfo{pages}{594--611}.
\newblock


\bibitem[Handa et~al\mbox{.}(2025)]%
        {economic}
\bibfield{author}{\bibinfo{person}{Kunal Handa}, \bibinfo{person}{Alex Tamkin}, \bibinfo{person}{Miles McCain}, \bibinfo{person}{Saffron Huang}, \bibinfo{person}{Esin Durmus}, \bibinfo{person}{Sarah Heck}, \bibinfo{person}{Jared Mueller}, \bibinfo{person}{Jerry Hong}, \bibinfo{person}{Stuart Ritchie}, \bibinfo{person}{Tim Belonax}, \bibinfo{person}{Kevin~K. Troy}, \bibinfo{person}{Dario Amodei}, \bibinfo{person}{Jared Kaplan}, \bibinfo{person}{Jack Clark}, {and} \bibinfo{person}{Deep Ganguli}.} \bibinfo{year}{2025}\natexlab{}.
\newblock \showarticletitle{Which Economic Tasks are Performed with {AI}? Evidence from Millions of {Claude} Conversations}.
\newblock \bibinfo{journal}{\emph{arXiv preprint arXiv:2503.04761}} (\bibinfo{year}{2025}).
\newblock


\bibitem[Heo et~al\mbox{.}(2024)]%
        {neupims}
\bibfield{author}{\bibinfo{person}{Guseul Heo}, \bibinfo{person}{Sangyeop Lee}, \bibinfo{person}{Jaehong Cho}, \bibinfo{person}{Hyunmin Choi}, \bibinfo{person}{Sanghyeon Lee}, \bibinfo{person}{Hyungkyu Ham}, \bibinfo{person}{Gwangsun Kim}, \bibinfo{person}{Divya Mahajan}, {and} \bibinfo{person}{Jongse Park}.} \bibinfo{year}{2024}\natexlab{}.
\newblock \showarticletitle{{NeuPIMs}: {NPU-PIM} Heterogeneous Acceleration for Batched {LLM} Inferencing}. In \bibinfo{booktitle}{\emph{Proceedings of the 29th ACM International Conference on Architectural Support for Programming Languages and Operating Systems, Volume 3}}. \bibinfo{pages}{722--737}.
\newblock


\bibitem[Hooper et~al\mbox{.}(2024)]%
        {kvquant}
\bibfield{author}{\bibinfo{person}{Coleman Hooper}, \bibinfo{person}{Sehoon Kim}, \bibinfo{person}{Hiva Mohammadzadeh}, \bibinfo{person}{Michael~W Mahoney}, \bibinfo{person}{Yakun~S Shao}, \bibinfo{person}{Kurt Keutzer}, {and} \bibinfo{person}{Amir Gholami}.} \bibinfo{year}{2024}\natexlab{}.
\newblock \showarticletitle{{KVQuant}: Towards 10 Million Context Length {LLM} Inference with {KV} Cache Quantization}.
\newblock \bibinfo{journal}{\emph{Advances in Neural Information Processing Systems}}  \bibinfo{volume}{37} (\bibinfo{year}{2024}), \bibinfo{pages}{1270--1303}.
\newblock


\bibitem[Huang et~al\mbox{.}(2026)]%
        {nosa}
\bibfield{author}{\bibinfo{person}{Yuxiang Huang}, \bibinfo{person}{Pengjie Wang}, \bibinfo{person}{Jicheng Han}, \bibinfo{person}{Weilin Zhao}, \bibinfo{person}{Zhou Su}, \bibinfo{person}{Ao Sun}, \bibinfo{person}{Hongya Lyu}, \bibinfo{person}{Hengyu Zhao}, \bibinfo{person}{Yudong Wang}, \bibinfo{person}{Chaojun Xiao}, \bibinfo{person}{Xu Han}, {and} \bibinfo{person}{Zhiyuan Liu}.} \bibinfo{year}{2026}\natexlab{}.
\newblock \showarticletitle{{NOSA}: Native and Offloadable Sparse Attention}. In \bibinfo{booktitle}{\emph{Proceedings of the 2026 Conference on Empirical Methods in Natural Language Processing}}. \bibinfo{pages}{6356--6381}.
\newblock


\bibitem[Kim et~al\mbox{.}(2025b)]%
        {scalablepnm}
\bibfield{author}{\bibinfo{person}{Dowon Kim}, \bibinfo{person}{MinJae Lee}, \bibinfo{person}{Janghyeon Kim}, \bibinfo{person}{HyuckSung Kwon}, \bibinfo{person}{Hyeonggyu Jeong}, \bibinfo{person}{Sang-Soo Park}, \bibinfo{person}{Minyong Yoon}, \bibinfo{person}{Si-Dong Roh}, \bibinfo{person}{Yongsuk Kwon}, \bibinfo{person}{Jinin So}, {and} \bibinfo{person}{Jungwook Choi}.} \bibinfo{year}{2025}\natexlab{b}.
\newblock \showarticletitle{Scalable Processing-Near-Memory for {1M}-Token {LLM} Inference: {CXL}-Enabled {KV}-Cache Management Beyond {GPU} Limits}. In \bibinfo{booktitle}{\emph{2025 34th International Conference on Parallel Architectures and Compilation Techniques (PACT)}}. IEEE, \bibinfo{pages}{1--13}.
\newblock


\bibitem[Kim and So(2026)]%
        {xcena_mx1}
\bibfield{author}{\bibinfo{person}{Harry Kim} {and} \bibinfo{person}{Jinin So}.} \bibinfo{year}{2026}\natexlab{}.
\newblock \showarticletitle{{XCENA} {MX1} {CXL} Computational Memory Device}. In \bibinfo{booktitle}{\emph{Hot Chips: A Symposium on High Performance Chips (HC38)}}.
\newblock


\bibitem[Kim et~al\mbox{.}(2025a)]%
        {pimba}
\bibfield{author}{\bibinfo{person}{Wonung Kim}, \bibinfo{person}{Yubin Lee}, \bibinfo{person}{Yoonsung Kim}, \bibinfo{person}{Jinwoo Hwang}, \bibinfo{person}{Seongryong Oh}, \bibinfo{person}{Jiyong Jung}, \bibinfo{person}{Aziz Huseynov}, \bibinfo{person}{Woong~Gyu Park}, \bibinfo{person}{Chang~Hyun Park}, \bibinfo{person}{Divya Mahajan}, {and} \bibinfo{person}{Jongse Park}.} \bibinfo{year}{2025}\natexlab{a}.
\newblock \showarticletitle{{Pimba}: A Processing-in-Memory Acceleration for Post-Transformer Large Language Model Serving}. In \bibinfo{booktitle}{\emph{Proceedings of the 58th IEEE/ACM International Symposium on Microarchitecture}}. \bibinfo{pages}{292--307}.
\newblock


\bibitem[Kwon et~al\mbox{.}(2026)]%
        {pimphony}
\bibfield{author}{\bibinfo{person}{Hyucksung Kwon}, \bibinfo{person}{Kyungmo Koo}, \bibinfo{person}{Janghyeon Kim}, \bibinfo{person}{Woongkyu Lee}, \bibinfo{person}{Minjae Lee}, \bibinfo{person}{Gyeonggeun Jung}, \bibinfo{person}{Hyungdeok Lee}, \bibinfo{person}{Yousub Jung}, \bibinfo{person}{Jaehan Park}, \bibinfo{person}{Yosub Song}, \bibinfo{person}{Byeongsu Yang}, \bibinfo{person}{Haerang Choi}, \bibinfo{person}{Guhyun Kim}, \bibinfo{person}{Jongsoon Won}, \bibinfo{person}{Woojae Shin}, \bibinfo{person}{Changhyun Kim}, \bibinfo{person}{Gyeongcheol Shin}, \bibinfo{person}{Yongkee Kwon}, \bibinfo{person}{Ilkon Kim}, \bibinfo{person}{Euicheol Lim}, \bibinfo{person}{John Kim}, {and} \bibinfo{person}{Jungwook Choi}.} \bibinfo{year}{2026}\natexlab{}.
\newblock \showarticletitle{{PIMphony}: Overcoming Bandwidth and Capacity Inefficiency in {PIM}-Based Long-Context {LLM} Inference System}. In \bibinfo{booktitle}{\emph{2026 IEEE International Symposium on High Performance Computer Architecture (HPCA)}}. IEEE, \bibinfo{pages}{1--21}.
\newblock


\bibitem[Kwon et~al\mbox{.}(2023)]%
        {vllm}
\bibfield{author}{\bibinfo{person}{Woosuk Kwon}, \bibinfo{person}{Zhuohan Li}, \bibinfo{person}{Siyuan Zhuang}, \bibinfo{person}{Ying Sheng}, \bibinfo{person}{Lianmin Zheng}, \bibinfo{person}{Cody~Hao Yu}, \bibinfo{person}{Joseph Gonzalez}, \bibinfo{person}{Hao Zhang}, {and} \bibinfo{person}{Ion Stoica}.} \bibinfo{year}{2023}\natexlab{}.
\newblock \showarticletitle{Efficient Memory Management for Large Language Model Serving with {PagedAttention}}. In \bibinfo{booktitle}{\emph{Proceedings of the 29th symposium on operating systems principles}}. \bibinfo{pages}{611--626}.
\newblock


\bibitem[Lai et~al\mbox{.}(2026)]%
        {minimax_m3}
\bibfield{author}{\bibinfo{person}{Xunhao Lai}, \bibinfo{person}{Weiqi Xu}, \bibinfo{person}{Yufeng Yang}, \bibinfo{person}{Qiaorui Chen}, \bibinfo{person}{Yang Xu}, \bibinfo{person}{Lunbin Zeng}, \bibinfo{person}{Xiaolong Li}, \bibinfo{person}{Haohai Sun}, \bibinfo{person}{Haichao Zhu}, \bibinfo{person}{Vito Zhang}, \bibinfo{person}{Jinkai Hu}, \bibinfo{person}{Jiayao Li}, \bibinfo{person}{Rui Gao}, \bibinfo{person}{Zekun Li}, \bibinfo{person}{Songquan Zhu}, \bibinfo{person}{Jingkai Zhou}, {and} \bibinfo{person}{Pengyu Zhao}.} \bibinfo{year}{2026}\natexlab{}.
\newblock \showarticletitle{{MiniMax} Sparse Attention}.
\newblock \bibinfo{journal}{\emph{arXiv preprint arXiv:2606.13392}} (\bibinfo{year}{2026}).
\newblock


\bibitem[Lee et~al\mbox{.}(2024)]%
        {infinigen}
\bibfield{author}{\bibinfo{person}{Wonbeom Lee}, \bibinfo{person}{Jungi Lee}, \bibinfo{person}{Junghwan Seo}, {and} \bibinfo{person}{Jaewoong Sim}.} \bibinfo{year}{2024}\natexlab{}.
\newblock \showarticletitle{{InfiniGen}: Efficient Generative Inference of Large Language Models with Dynamic {KV} Cache Management}. In \bibinfo{booktitle}{\emph{18th USENIX symposium on operating systems design and implementation (OSDI 24)}}. \bibinfo{pages}{155--172}.
\newblock


\bibitem[Li et~al\mbox{.}(2023)]%
        {pond}
\bibfield{author}{\bibinfo{person}{Huaicheng Li}, \bibinfo{person}{Daniel~S. Berger}, \bibinfo{person}{Lisa Hsu}, \bibinfo{person}{Daniel Ernst}, \bibinfo{person}{Pantea Zardoshti}, \bibinfo{person}{Stanko Novakovic}, \bibinfo{person}{Monish Shah}, \bibinfo{person}{Samir Rajadnya}, \bibinfo{person}{Scott Lee}, \bibinfo{person}{Ishwar Agarwal}, \bibinfo{person}{Mark~D. Hill}, \bibinfo{person}{Marcus Fontoura}, {and} \bibinfo{person}{Ricardo Bianchini}.} \bibinfo{year}{2023}\natexlab{}.
\newblock \showarticletitle{{Pond}: {CXL}-Based Memory Pooling Systems for Cloud Platforms}. In \bibinfo{booktitle}{\emph{Proceedings of the 28th ACM International Conference on Architectural Support for Programming Languages and Operating Systems, Volume 2}}. \bibinfo{pages}{574--587}.
\newblock


\bibitem[Li et~al\mbox{.}(2024)]%
        {snapkv}
\bibfield{author}{\bibinfo{person}{Yuhong Li}, \bibinfo{person}{Yingbing Huang}, \bibinfo{person}{Bowen Yang}, \bibinfo{person}{Bharat Venkitesh}, \bibinfo{person}{Acyr Locatelli}, \bibinfo{person}{Hanchen Ye}, \bibinfo{person}{Tianle Cai}, \bibinfo{person}{Patrick Lewis}, {and} \bibinfo{person}{Deming Chen}.} \bibinfo{year}{2024}\natexlab{}.
\newblock \showarticletitle{{SnapKV}: {LLM} Knows What You are Looking for Before Generation}.
\newblock \bibinfo{journal}{\emph{Advances in Neural Information Processing Systems}}  \bibinfo{volume}{37} (\bibinfo{year}{2024}), \bibinfo{pages}{22947--22970}.
\newblock


\bibitem[Li et~al\mbox{.}(2026a)]%
        {openresearcher}
\bibfield{author}{\bibinfo{person}{Zhuofeng Li}, \bibinfo{person}{Dongfu Jiang}, \bibinfo{person}{Xueguang Ma}, \bibinfo{person}{Haoxiang Zhang}, \bibinfo{person}{Ping Nie}, \bibinfo{person}{Yuyu Zhang}, \bibinfo{person}{Kai Zou}, \bibinfo{person}{Jianwen Xie}, \bibinfo{person}{Yu Zhang}, {and} \bibinfo{person}{Wenhu Chen}.} \bibinfo{year}{2026}\natexlab{a}.
\newblock \showarticletitle{{OpenResearcher}: A Fully Open Pipeline for Long-Horizon Deep Research Trajectory Synthesis}. In \bibinfo{booktitle}{\emph{Proceedings of the 2026 Conference on Empirical Methods in Natural Language Processing}}. \bibinfo{pages}{908--937}.
\newblock


\bibitem[Li et~al\mbox{.}(2026b)]%
        {sas}
\bibfield{author}{\bibinfo{person}{Zhiwei Li}, \bibinfo{person}{Lei Zhu}, \bibinfo{person}{Hao Gu}, \bibinfo{person}{Xiang Hu}, \bibinfo{person}{Yan Wang}, \bibinfo{person}{Haitao Mi}, \bibinfo{person}{Sirui Han}, \bibinfo{person}{Leo Liang}, {and} \bibinfo{person}{Zhijiang Guo}.} \bibinfo{year}{2026}\natexlab{b}.
\newblock \showarticletitle{SAS: Simple Attention Sparsification via End-to-End Optimization of Context Ranking}.
\newblock \bibinfo{journal}{\emph{arXiv preprint arXiv:2609.13141}} (\bibinfo{year}{2026}).
\newblock


\bibitem[Liu et~al\mbox{.}(2025)]%
        {retrievalattention}
\bibfield{author}{\bibinfo{person}{Di Liu}, \bibinfo{person}{Meng Chen}, \bibinfo{person}{Baotong Lu}, \bibinfo{person}{Huiqiang Jiang}, \bibinfo{person}{Zhenhua Han}, \bibinfo{person}{Qianxi Zhang}, \bibinfo{person}{Qi Chen}, \bibinfo{person}{Chengruidong Zhang}, \bibinfo{person}{Bailu Ding}, \bibinfo{person}{Kai Zhang}, \bibinfo{person}{Chen Chen}, \bibinfo{person}{Fan Yang}, \bibinfo{person}{Yuqing Yang}, {and} \bibinfo{person}{Lili Qiu}.} \bibinfo{year}{2025}\natexlab{}.
\newblock \showarticletitle{{RetrievalAttention}: Accelerating Long-Context {LLM} Inference via Vector Retrieval}.
\newblock \bibinfo{journal}{\emph{Advances in Neural Information Processing Systems}}  \bibinfo{volume}{38} (\bibinfo{year}{2025}), \bibinfo{pages}{54358--54385}.
\newblock


\bibitem[Liu et~al\mbox{.}(2026a)]%
        {echo}
\bibfield{author}{\bibinfo{person}{Guangda Liu}, \bibinfo{person}{Wenhao Chen}, \bibinfo{person}{Chengwei Li}, \bibinfo{person}{Zhenyu Ning}, \bibinfo{person}{Jing Lin}, \bibinfo{person}{Yiwu Yao}, \bibinfo{person}{Quan Chen}, \bibinfo{person}{Shixuan Sun}, \bibinfo{person}{Jieru Zhao}, {and} \bibinfo{person}{Minyi Guo}.} \bibinfo{year}{2026}\natexlab{a}.
\newblock \showarticletitle{{ECHO}: Efficient {KV} Cache Offloading with Lossless Prefetching for Serving Native Sparse Attention {LLMs}}. In \bibinfo{booktitle}{\emph{20th USENIX Symposium on Operating Systems Design and Implementation (OSDI 26)}}. \bibinfo{pages}{17--37}.
\newblock


\bibitem[Liu et~al\mbox{.}(2026b)]%
        {l3}
\bibfield{author}{\bibinfo{person}{Qingyuan Liu}, \bibinfo{person}{Liyan Chen}, \bibinfo{person}{Haocheng Wang}, \bibinfo{person}{Yanning Yang}, \bibinfo{person}{Dong Du}, \bibinfo{person}{Zhigang Mao}, \bibinfo{person}{Naifeng Jing}, \bibinfo{person}{Yubin Xia}, {and} \bibinfo{person}{Haibo Chen}.} \bibinfo{year}{2026}\natexlab{b}.
\newblock \showarticletitle{{CHIME}: A Case for Efficient Long-Context Attention-{FC} Disaggregated Inference with {DIMM-PIM}}. In \bibinfo{booktitle}{\emph{2026 ACM/IEEE 53rd Annual International Symposium on Computer Architecture (ISCA)}}. \bibinfo{pages}{1--17}.
\newblock


\bibitem[Liu et~al\mbox{.}(2024)]%
        {kivi}
\bibfield{author}{\bibinfo{person}{Zirui Liu}, \bibinfo{person}{Jiayi Yuan}, \bibinfo{person}{Hongye Jin}, \bibinfo{person}{Shaochen Zhong}, \bibinfo{person}{Zhaozhuo Xu}, \bibinfo{person}{Vladimir Braverman}, \bibinfo{person}{Beidi Chen}, {and} \bibinfo{person}{Xia Hu}.} \bibinfo{year}{2024}\natexlab{}.
\newblock \showarticletitle{{KIVI}: A Tuning-Free Asymmetric 2bit Quantization for {KV} Cache}. In \bibinfo{booktitle}{\emph{Proceedings of the 41st International Conference on Machine Learning}} \emph{(\bibinfo{series}{Proceedings of Machine Learning Research}, Vol.~\bibinfo{volume}{235})}. \bibinfo{publisher}{PMLR}, \bibinfo{pages}{32332--32344}.
\newblock


\bibitem[Lu et~al\mbox{.}(2025)]%
        {moba}
\bibfield{author}{\bibinfo{person}{Enzhe Lu}, \bibinfo{person}{Zhejun Jiang}, \bibinfo{person}{Jingyuan Liu}, \bibinfo{person}{Yulun Du}, \bibinfo{person}{Tao Jiang}, \bibinfo{person}{Chao Hong}, \bibinfo{person}{Shaowei Liu}, \bibinfo{person}{Weiran He}, \bibinfo{person}{Enming Yuan}, \bibinfo{person}{Yuzhi Wang}, \bibinfo{person}{Zhiqi Huang}, \bibinfo{person}{Huan Yuan}, \bibinfo{person}{Suting Xu}, \bibinfo{person}{Xinran Xu}, \bibinfo{person}{Guokun Lai}, \bibinfo{person}{Yanru Chen}, \bibinfo{person}{Huabin Zheng}, \bibinfo{person}{Junjie Yan}, \bibinfo{person}{Jianlin Su}, \bibinfo{person}{Yuxin Wu}, \bibinfo{person}{Yutao Zhang}, \bibinfo{person}{Zhilin Yang}, \bibinfo{person}{Xinyu Zhou}, \bibinfo{person}{Mingxing Zhang}, {and} \bibinfo{person}{Jiezhong Qiu}.} \bibinfo{year}{2025}\natexlab{}.
\newblock \showarticletitle{{MoBA}: Mixture of Block Attention for Long-Context {LLMs}}.
\newblock \bibinfo{journal}{\emph{Advances in Neural Information Processing Systems}}  \bibinfo{volume}{38} (\bibinfo{year}{2025}), \bibinfo{pages}{17790--17815}.
\newblock


\bibitem[Malik and Park(2026)]%
        {marvell_structera_sk_hynix_2026}
\bibfield{author}{\bibinfo{person}{Khurram Malik} {and} \bibinfo{person}{Kangkyu Park}.} \bibinfo{year}{2026}\natexlab{}.
\newblock \bibinfo{title}{Accelerating AI Infrastructure with Marvell Structera A and SK hynix CXL Memory: Enabling Efficient Near-Memory Processing}.
\newblock \bibinfo{howpublished}{Marvell Blog}.
\newblock
\urldef\tempurl%
\url{https://www.marvell.com/blogs/accelerating-ai-infrastructure-marvell-structera-sk-hynix-cxl-memory.html}
\showURL{%
\tempurl}
\newblock
\shownote{Accessed: 2026-11-07}.


\bibitem[Maruf et~al\mbox{.}(2023)]%
        {tpp}
\bibfield{author}{\bibinfo{person}{Hasan~Al Maruf}, \bibinfo{person}{Hao Wang}, \bibinfo{person}{Abhishek Dhanotia}, \bibinfo{person}{Johannes Weiner}, \bibinfo{person}{Niket Agarwal}, \bibinfo{person}{Pallab Bhattacharya}, \bibinfo{person}{Chris Petersen}, \bibinfo{person}{Mosharaf Chowdhury}, \bibinfo{person}{Shobhit Kanaujia}, {and} \bibinfo{person}{Prakash Chauhan}.} \bibinfo{year}{2023}\natexlab{}.
\newblock \showarticletitle{{TPP}: Transparent Page Placement for {CXL}-Enabled Tiered-Memory}. In \bibinfo{booktitle}{\emph{Proceedings of the 28th ACM International Conference on Architectural Support for Programming Languages and Operating Systems, Volume 3}}. \bibinfo{pages}{742--755}.
\newblock


\bibitem[{Marvell Technology}(2024)]%
        {marvell_structera_a}
\bibfield{author}{\bibinfo{person}{{Marvell Technology}}.} \bibinfo{year}{2024}\natexlab{}.
\newblock \bibinfo{title}{Structera {A} 2504 Memory-Expansion Controller: {CXL} 2.0 {DDR5} 4-Channel Accelerator}.
\newblock \bibinfo{howpublished}{Product Brief}.
\newblock
\urldef\tempurl%
\url{https://jp.marvell.com/content/dam/marvell/en/public-collateral/assets/marvell-structera-a-2504-near-memory-accelerator-product-brief.pdf}
\showURL{%
\tempurl}
\newblock
\shownote{Accessed: 2026-11-10}.


\bibitem[{NVIDIA}(2023)]%
        {h100nvl}
\bibfield{author}{\bibinfo{person}{{NVIDIA}}.} \bibinfo{year}{2023}\natexlab{}.
\newblock \bibinfo{title}{{NVIDIA} {H100} {NVL} {GPU} Datasheet}.
\newblock \bibinfo{howpublished}{Datasheet}.
\newblock
\urldef\tempurl%
\url{https://www.pny.com/File%20Library/Company/Support/Product%20Brochures/NVIDIA%20Data%20Center%20GPUs/english/H100-NVL-Datasheet.pdf}
\showURL{%
\tempurl}
\newblock
\shownote{Accessed: 2026-11-10}.


\bibitem[Park et~al\mbox{.}(2024a)]%
        {attacc}
\bibfield{author}{\bibinfo{person}{Jaehyun Park}, \bibinfo{person}{Jaewan Choi}, \bibinfo{person}{Kwanhee Kyung}, \bibinfo{person}{Michael~Jaemin Kim}, \bibinfo{person}{Yongsuk Kwon}, \bibinfo{person}{Nam~Sung Kim}, {and} \bibinfo{person}{Jung~Ho Ahn}.} \bibinfo{year}{2024}\natexlab{a}.
\newblock \showarticletitle{{AttAcc}! Unleashing the Power of {PIM} for Batched Transformer-Based Generative Model Inference}. In \bibinfo{booktitle}{\emph{Proceedings of the 29th ACM International Conference on Architectural Support for Programming Languages and Operating Systems, Volume 2}}. \bibinfo{pages}{103--119}.
\newblock


\bibitem[Park et~al\mbox{.}(2024b)]%
        {cxlpnm}
\bibfield{author}{\bibinfo{person}{Sang-Soo Park}, \bibinfo{person}{KyungSoo Kim}, \bibinfo{person}{Jinin So}, \bibinfo{person}{Jin Jung}, \bibinfo{person}{Jonggeon Lee}, \bibinfo{person}{Kyoungwan Woo}, \bibinfo{person}{Nayeon Kim}, \bibinfo{person}{Younghyun Lee}, \bibinfo{person}{Hyungyo Kim}, \bibinfo{person}{Yongsuk Kwon}, \bibinfo{person}{Jinhyun Kim}, \bibinfo{person}{Jieun Lee}, \bibinfo{person}{YeonGon Cho}, \bibinfo{person}{Yongmin Tai}, \bibinfo{person}{Jeonghyeon Cho}, \bibinfo{person}{Hoyoung Song}, \bibinfo{person}{Jung~Ho Ahn}, {and} \bibinfo{person}{Nam~Sung Kim}.} \bibinfo{year}{2024}\natexlab{b}.
\newblock \showarticletitle{An {LPDDR}-Based {CXL-PNM} Platform for {TCO}-Efficient Inference of Transformer-Based Large Language Models}. In \bibinfo{booktitle}{\emph{2024 IEEE International Symposium on High-Performance Computer Architecture (HPCA)}}. IEEE, \bibinfo{pages}{970--982}.
\newblock


\bibitem[Patel et~al\mbox{.}(2024)]%
        {splitwise}
\bibfield{author}{\bibinfo{person}{Pratyush Patel}, \bibinfo{person}{Esha Choukse}, \bibinfo{person}{Chaojie Zhang}, \bibinfo{person}{Aashaka Shah}, \bibinfo{person}{{\'I}{\~n}igo Goiri}, \bibinfo{person}{Saeed Maleki}, {and} \bibinfo{person}{Ricardo Bianchini}.} \bibinfo{year}{2024}\natexlab{}.
\newblock \showarticletitle{{Splitwise}: Efficient Generative {LLM} Inference Using Phase Splitting}. In \bibinfo{booktitle}{\emph{2024 ACM/IEEE 51st Annual International Symposium on Computer Architecture (ISCA)}}. IEEE, \bibinfo{pages}{118--132}.
\newblock


\bibitem[Pope et~al\mbox{.}(2023)]%
        {pope2023scaling}
\bibfield{author}{\bibinfo{person}{Reiner Pope}, \bibinfo{person}{Sholto Douglas}, \bibinfo{person}{Aakanksha Chowdhery}, \bibinfo{person}{Jacob Devlin}, \bibinfo{person}{James Bradbury}, \bibinfo{person}{Jonathan Heek}, \bibinfo{person}{Kefan Xiao}, \bibinfo{person}{Shivani Agrawal}, {and} \bibinfo{person}{Jeff Dean}.} \bibinfo{year}{2023}\natexlab{}.
\newblock \showarticletitle{Efficiently Scaling Transformer Inference}.
\newblock \bibinfo{journal}{\emph{Proceedings of machine learning and systems}}  \bibinfo{volume}{5} (\bibinfo{year}{2023}), \bibinfo{pages}{606--624}.
\newblock


\bibitem[Qin et~al\mbox{.}(2025)]%
        {mooncake}
\bibfield{author}{\bibinfo{person}{Ruoyu Qin}, \bibinfo{person}{Zheming Li}, \bibinfo{person}{Weiran He}, \bibinfo{person}{Jialei Cui}, \bibinfo{person}{Feng Ren}, \bibinfo{person}{Mingxing Zhang}, \bibinfo{person}{Yongwei Wu}, \bibinfo{person}{Weimin Zheng}, {and} \bibinfo{person}{Xinran Xu}.} \bibinfo{year}{2025}\natexlab{}.
\newblock \showarticletitle{{Mooncake}: Trading More Storage for Less Computation --- A {KVCache}-Centric Architecture for Serving {LLM} Chatbot}. In \bibinfo{booktitle}{\emph{23rd USENIX conference on file and storage technologies (FAST 25)}}. \bibinfo{pages}{155--170}.
\newblock


\bibitem[Quinn et~al\mbox{.}(2025)]%
        {longsight}
\bibfield{author}{\bibinfo{person}{Derrick Quinn}, \bibinfo{person}{E~Ezgi Y{\"u}cel}, \bibinfo{person}{Jinkwon Kim}, \bibinfo{person}{Jos{\'e}~F Mart{\'\i}nez}, {and} \bibinfo{person}{Mohammad Alian}.} \bibinfo{year}{2025}\natexlab{}.
\newblock \showarticletitle{{LongSight}: Compute-Enabled Memory to Accelerate Large-Context {LLMs} via Sparse Attention}. In \bibinfo{booktitle}{\emph{Proceedings of the 58th IEEE/ACM International Symposium on Microarchitecture}}. \bibinfo{pages}{34--48}.
\newblock


\bibitem[Sarangi and Baas(2021)]%
        {deepscaletool}
\bibfield{author}{\bibinfo{person}{Satyabrata Sarangi} {and} \bibinfo{person}{Bevan Baas}.} \bibinfo{year}{2021}\natexlab{}.
\newblock \showarticletitle{{DeepScaleTool}: A Tool for the Accurate Estimation of Technology Scaling in the Deep-Submicron Era}. In \bibinfo{booktitle}{\emph{2021 IEEE International Symposium on Circuits and Systems (ISCAS)}}. IEEE, \bibinfo{pages}{1--5}.
\newblock


\bibitem[Sheng et~al\mbox{.}(2023)]%
        {flexgen}
\bibfield{author}{\bibinfo{person}{Ying Sheng}, \bibinfo{person}{Lianmin Zheng}, \bibinfo{person}{Binhang Yuan}, \bibinfo{person}{Zhuohan Li}, \bibinfo{person}{Max Ryabinin}, \bibinfo{person}{Beidi Chen}, \bibinfo{person}{Percy Liang}, \bibinfo{person}{Christopher R{\'e}}, \bibinfo{person}{Ion Stoica}, {and} \bibinfo{person}{Ce Zhang}.} \bibinfo{year}{2023}\natexlab{}.
\newblock \showarticletitle{{FlexGen}: High-Throughput Generative Inference of Large Language Models with a Single {GPU}}. In \bibinfo{booktitle}{\emph{International Conference on Machine Learning}}. PMLR, \bibinfo{pages}{31094--31116}.
\newblock


\bibitem[Shinn et~al\mbox{.}(2023)]%
        {reflexion}
\bibfield{author}{\bibinfo{person}{Noah Shinn}, \bibinfo{person}{Federico Cassano}, \bibinfo{person}{Ashwin Gopinath}, \bibinfo{person}{Karthik Narasimhan}, {and} \bibinfo{person}{Shunyu Yao}.} \bibinfo{year}{2023}\natexlab{}.
\newblock \showarticletitle{{Reflexion}: Language Agents with Verbal Reinforcement Learning}.
\newblock \bibinfo{journal}{\emph{Advances in neural information processing systems}}  \bibinfo{volume}{36} (\bibinfo{year}{2023}), \bibinfo{pages}{8634--8652}.
\newblock


\bibitem[Sim et~al\mbox{.}(2023)]%
        {cms}
\bibfield{author}{\bibinfo{person}{Joonseop Sim}, \bibinfo{person}{Soohong Ahn}, \bibinfo{person}{Taeyoung Ahn}, \bibinfo{person}{Seungyong Lee}, \bibinfo{person}{Myunghyun Rhee}, \bibinfo{person}{Jooyoung Kim}, \bibinfo{person}{Kwangsik Shin}, \bibinfo{person}{Donguk Moon}, \bibinfo{person}{Euiseok Kim}, {and} \bibinfo{person}{Kyoung Park}.} \bibinfo{year}{2023}\natexlab{}.
\newblock \showarticletitle{Computational {CXL}-Memory Solution for Accelerating Memory-Intensive Applications}.
\newblock \bibinfo{journal}{\emph{IEEE Computer Architecture Letters}} \bibinfo{volume}{22}, \bibinfo{number}{1} (\bibinfo{year}{2023}), \bibinfo{pages}{5--8}.
\newblock


\bibitem[Sun et~al\mbox{.}(2025)]%
        {shadowkv}
\bibfield{author}{\bibinfo{person}{Hanshi Sun}, \bibinfo{person}{Li-Wen Chang}, \bibinfo{person}{Wenlei Bao}, \bibinfo{person}{Size Zheng}, \bibinfo{person}{Ningxin Zheng}, \bibinfo{person}{Xin Liu}, \bibinfo{person}{Harry Dong}, \bibinfo{person}{Yuejie Chi}, {and} \bibinfo{person}{Beidi Chen}.} \bibinfo{year}{2025}\natexlab{}.
\newblock \showarticletitle{{S}hadow{KV}: {KV} Cache in Shadows for High-Throughput Long-Context {LLM} Inference}. In \bibinfo{booktitle}{\emph{Proceedings of the 42nd International Conference on Machine Learning}} \emph{(\bibinfo{series}{Proceedings of Machine Learning Research}, Vol.~\bibinfo{volume}{267})}. \bibinfo{publisher}{PMLR}, \bibinfo{pages}{57355--57373}.
\newblock


\bibitem[Tang et~al\mbox{.}(2024)]%
        {quest}
\bibfield{author}{\bibinfo{person}{Jiaming Tang}, \bibinfo{person}{Yilong Zhao}, \bibinfo{person}{Kan Zhu}, \bibinfo{person}{Guangxuan Xiao}, \bibinfo{person}{Baris Kasikci}, {and} \bibinfo{person}{Song Han}.} \bibinfo{year}{2024}\natexlab{}.
\newblock \showarticletitle{{QUEST}: Query-Aware Sparsity for Efficient Long-Context {LLM} Inference}. In \bibinfo{booktitle}{\emph{Proceedings of the 41st International Conference on Machine Learning}} \emph{(\bibinfo{series}{Proceedings of Machine Learning Research}, Vol.~\bibinfo{volume}{235})}. \bibinfo{publisher}{PMLR}, \bibinfo{pages}{47901--47911}.
\newblock


\bibitem[Wang et~al\mbox{.}(2025)]%
        {openhands}
\bibfield{author}{\bibinfo{person}{Xingyao Wang}, \bibinfo{person}{Boxuan Li}, \bibinfo{person}{Yufan Song}, \bibinfo{person}{Frank~F. Xu}, \bibinfo{person}{Xiangru Tang}, \bibinfo{person}{Mingchen Zhuge}, \bibinfo{person}{Jiayi Pan}, \bibinfo{person}{Yueqi Song}, \bibinfo{person}{Bowen Li}, \bibinfo{person}{Jaskirat Singh}, \bibinfo{person}{Hoang~H. Tran}, \bibinfo{person}{Fuqiang Li}, \bibinfo{person}{Ren Ma}, \bibinfo{person}{Mingzhang Zheng}, \bibinfo{person}{Bill Qian}, \bibinfo{person}{Yanjun Shao}, \bibinfo{person}{Niklas Muennighoff}, \bibinfo{person}{Yizhe Zhang}, \bibinfo{person}{Binyuan Hui}, \bibinfo{person}{Junyang Lin}, \bibinfo{person}{Robert Brennan}, \bibinfo{person}{Hao Peng}, \bibinfo{person}{Heng Ji}, {and} \bibinfo{person}{Graham Neubig}.} \bibinfo{year}{2025}\natexlab{}.
\newblock \showarticletitle{{OpenHands}: An Open Platform for {AI} Software Developers as Generalist Agents}. In \bibinfo{booktitle}{\emph{International Conference on Learning Representations}}, Vol.~\bibinfo{volume}{2025}. \bibinfo{pages}{65882--65919}.
\newblock


\bibitem[Xiao et~al\mbox{.}(2024b)]%
        {infllm}
\bibfield{author}{\bibinfo{person}{Chaojun Xiao}, \bibinfo{person}{Pengle Zhang}, \bibinfo{person}{Xu Han}, \bibinfo{person}{Guangxuan Xiao}, \bibinfo{person}{Yankai Lin}, \bibinfo{person}{Zhengyan Zhang}, \bibinfo{person}{Zhiyuan Liu}, {and} \bibinfo{person}{Maosong Sun}.} \bibinfo{year}{2024}\natexlab{b}.
\newblock \showarticletitle{{InfLLM}: Training-Free Long-Context Extrapolation for {LLMs} with an Efficient Context Memory}.
\newblock \bibinfo{journal}{\emph{Advances in neural information processing systems}}  \bibinfo{volume}{37} (\bibinfo{year}{2024}), \bibinfo{pages}{119638--119661}.
\newblock


\bibitem[Xiao et~al\mbox{.}(2024a)]%
        {streamingllm}
\bibfield{author}{\bibinfo{person}{Guangxuan Xiao}, \bibinfo{person}{Yuandong Tian}, \bibinfo{person}{Beidi Chen}, \bibinfo{person}{Song Han}, {and} \bibinfo{person}{Mike Lewis}.} \bibinfo{year}{2024}\natexlab{a}.
\newblock \showarticletitle{Efficient Streaming Language Models with Attention Sinks}. In \bibinfo{booktitle}{\emph{International Conference on Learning Representations}}, Vol.~\bibinfo{volume}{2024}. \bibinfo{pages}{21875--21895}.
\newblock


\bibitem[Xie et~al\mbox{.}(2026a)]%
        {hisparse}
\bibfield{author}{\bibinfo{person}{Zhiqiang Xie}, \bibinfo{person}{Zhangheng Huang}, \bibinfo{person}{Tingwei Huang}, \bibinfo{person}{Ziyi Xu}, \bibinfo{person}{Ruiyang Ma}, {and} \bibinfo{person}{Christos Kozyrakis}.} \bibinfo{year}{2026}\natexlab{a}.
\newblock \showarticletitle{{HiSparse}: Scaling Sparse-Attention Decoding with Hierarchical {KV} Cache Management}.
\newblock \bibinfo{journal}{\emph{arXiv preprint arXiv:2608.07009}} (\bibinfo{year}{2026}).
\newblock


\bibitem[Xie et~al\mbox{.}(2026b)]%
        {strata}
\bibfield{author}{\bibinfo{person}{Zhiqiang Xie}, \bibinfo{person}{Ziyi Xu}, \bibinfo{person}{Mark Zhao}, \bibinfo{person}{Yuwei An}, \bibinfo{person}{Vikram~Sharma Mailthody}, \bibinfo{person}{Scott Mahlke}, \bibinfo{person}{Michael Garland}, {and} \bibinfo{person}{Christos Kozyrakis}.} \bibinfo{year}{2026}\natexlab{b}.
\newblock \showarticletitle{{Strata}: Hierarchical Context Caching for Long Context Language Model Serving}. In \bibinfo{booktitle}{\emph{20th USENIX Symposium on Operating Systems Design and Implementation (OSDI 26)}}. \bibinfo{pages}{1--16}.
\newblock


\bibitem[Yang et~al\mbox{.}(2025)]%
        {qwen3}
\bibfield{author}{\bibinfo{person}{An Yang}, \bibinfo{person}{Anfeng Li}, \bibinfo{person}{Baosong Yang}, \bibinfo{person}{Beichen Zhang}, \bibinfo{person}{Binyuan Hui}, \bibinfo{person}{Bo Zheng}, \bibinfo{person}{Bowen Yu}, \bibinfo{person}{Chang Gao}, \bibinfo{person}{Chengen Huang}, \bibinfo{person}{Chenxu Lv}, \bibinfo{person}{Chujie Zheng}, \bibinfo{person}{Dayiheng Liu}, \bibinfo{person}{Fan Zhou}, \bibinfo{person}{Fei Huang}, \bibinfo{person}{Feng Hu}, \bibinfo{person}{Hao Ge}, \bibinfo{person}{Haoran Wei}, \bibinfo{person}{Huan Lin}, \bibinfo{person}{Jialong Tang}, \bibinfo{person}{Jian Yang}, \bibinfo{person}{Jianhong Tu}, \bibinfo{person}{Jianwei Zhang}, \bibinfo{person}{Jianxin Yang}, \bibinfo{person}{Jiaxi Yang}, \bibinfo{person}{Jing Zhou}, \bibinfo{person}{Jingren Zhou}, \bibinfo{person}{Junyang Lin}, \bibinfo{person}{Kai Dang}, \bibinfo{person}{Keqin Bao}, \bibinfo{person}{Kexin Yang}, \bibinfo{person}{Le Yu}, \bibinfo{person}{Lianghao Deng}, \bibinfo{person}{Mei Li}, \bibinfo{person}{Mingfeng
  Xue}, \bibinfo{person}{Mingze Li}, \bibinfo{person}{Pei Zhang}, \bibinfo{person}{Peng Wang}, \bibinfo{person}{Qin Zhu}, \bibinfo{person}{Rui Men}, \bibinfo{person}{Ruize Gao}, \bibinfo{person}{Shixuan Liu}, \bibinfo{person}{Shuang Luo}, \bibinfo{person}{Tianhao Li}, \bibinfo{person}{Tianyi Tang}, \bibinfo{person}{Wenbiao Yin}, \bibinfo{person}{Xingzhang Ren}, \bibinfo{person}{Xinyu Wang}, \bibinfo{person}{Xinyu Zhang}, \bibinfo{person}{Xuancheng Ren}, \bibinfo{person}{Yang Fan}, \bibinfo{person}{Yang Su}, \bibinfo{person}{Yichang Zhang}, \bibinfo{person}{Yinger Zhang}, \bibinfo{person}{Yu Wan}, \bibinfo{person}{Yuqiong Liu}, \bibinfo{person}{Zekun Wang}, \bibinfo{person}{Zeyu Cui}, \bibinfo{person}{Zhenru Zhang}, \bibinfo{person}{Zhipeng Zhou}, {and} \bibinfo{person}{Zihan Qiu}.} \bibinfo{year}{2025}\natexlab{}.
\newblock \showarticletitle{{Qwen3} Technical Report}.
\newblock \bibinfo{journal}{\emph{arXiv preprint arXiv:2505.09388}} (\bibinfo{year}{2025}).
\newblock


\bibitem[Yang et~al\mbox{.}(2024)]%
        {sweagent}
\bibfield{author}{\bibinfo{person}{John Yang}, \bibinfo{person}{Carlos Jimenez}, \bibinfo{person}{Alexander Wettig}, \bibinfo{person}{Kilian Lieret}, \bibinfo{person}{Shunyu Yao}, \bibinfo{person}{Karthik Narasimhan}, {and} \bibinfo{person}{Ofir Press}.} \bibinfo{year}{2024}\natexlab{}.
\newblock \showarticletitle{{SWE-agent}: Agent-Computer Interfaces Enable Automated Software Engineering}.
\newblock \bibinfo{journal}{\emph{Advances in Neural Information Processing Systems}}  \bibinfo{volume}{37} (\bibinfo{year}{2024}), \bibinfo{pages}{50528--50652}.
\newblock


\bibitem[Yao et~al\mbox{.}(2023)]%
        {react}
\bibfield{author}{\bibinfo{person}{Shunyu Yao}, \bibinfo{person}{Jeffrey Zhao}, \bibinfo{person}{Dian Yu}, \bibinfo{person}{Nan Du}, \bibinfo{person}{Izhak Shafran}, \bibinfo{person}{Karthik~R Narasimhan}, {and} \bibinfo{person}{Yuan Cao}.} \bibinfo{year}{2023}\natexlab{}.
\newblock \showarticletitle{{ReAct}: Synergizing Reasoning and Acting in Language Models}. In \bibinfo{booktitle}{\emph{International Conference on Learning Representations}}.
\newblock


\bibitem[Yuan et~al\mbox{.}(2025)]%
        {nsa}
\bibfield{author}{\bibinfo{person}{Jingyang Yuan}, \bibinfo{person}{Huazuo Gao}, \bibinfo{person}{Damai Dai}, \bibinfo{person}{Junyu Luo}, \bibinfo{person}{Liang Zhao}, \bibinfo{person}{Zhengyan Zhang}, \bibinfo{person}{Zhenda Xie}, \bibinfo{person}{Yuxing Wei}, \bibinfo{person}{Lean Wang}, \bibinfo{person}{Zhiping Xiao}, \bibinfo{person}{Yuqing Wang}, \bibinfo{person}{Chong Ruan}, \bibinfo{person}{Ming Zhang}, \bibinfo{person}{Wenfeng Liang}, {and} \bibinfo{person}{Wangding Zeng}.} \bibinfo{year}{2025}\natexlab{}.
\newblock \showarticletitle{Native Sparse Attention: Hardware-Aligned and Natively Trainable Sparse Attention}. In \bibinfo{booktitle}{\emph{Proceedings of the 63rd Annual Meeting of the Association for Computational Linguistics (Volume 1: Long Papers)}}. \bibinfo{pages}{23078--23097}.
\newblock


\bibitem[Zhang et~al\mbox{.}(2025)]%
        {pqcache}
\bibfield{author}{\bibinfo{person}{Hailin Zhang}, \bibinfo{person}{Xiaodong Ji}, \bibinfo{person}{Yilin Chen}, \bibinfo{person}{Fangcheng Fu}, \bibinfo{person}{Xupeng Miao}, \bibinfo{person}{Xiaonan Nie}, \bibinfo{person}{Weipeng Chen}, {and} \bibinfo{person}{Bin Cui}.} \bibinfo{year}{2025}\natexlab{}.
\newblock \showarticletitle{{PQCache}: Product Quantization-Based {KVCache} for Long Context {LLM} Inference}.
\newblock \bibinfo{journal}{\emph{Proceedings of the ACM on Management of Data}} \bibinfo{volume}{3}, \bibinfo{number}{3} (\bibinfo{year}{2025}), \bibinfo{pages}{1--30}.
\newblock


\bibitem[Zhang et~al\mbox{.}(2026)]%
        {pulseinfer}
\bibfield{author}{\bibinfo{person}{Qiuyang Zhang}, \bibinfo{person}{Kai Zhou}, \bibinfo{person}{Kai Lu}, \bibinfo{person}{Haocheng Lu}, \bibinfo{person}{Jian Zhou}, \bibinfo{person}{Yuanpeng Su}, \bibinfo{person}{Kun Bao}, \bibinfo{person}{Jiguang Wan}, {and} \bibinfo{person}{Fei Wu}.} \bibinfo{year}{2026}\natexlab{}.
\newblock \showarticletitle{PulseInfer: I/O-Centric Sparse KV Cache Offloading for Efficient Long-Context LLM Decoding}.
\newblock \bibinfo{journal}{\emph{arXiv preprint arXiv:2609.34555}} (\bibinfo{year}{2026}).
\newblock


\bibitem[Zhang et~al\mbox{.}(2023)]%
        {h2o}
\bibfield{author}{\bibinfo{person}{Zhenyu Zhang}, \bibinfo{person}{Ying Sheng}, \bibinfo{person}{Tianyi Zhou}, \bibinfo{person}{Tianlong Chen}, \bibinfo{person}{Lianmin Zheng}, \bibinfo{person}{Ruisi Cai}, \bibinfo{person}{Zhao Song}, \bibinfo{person}{Yuandong Tian}, \bibinfo{person}{Christopher R{\'e}}, \bibinfo{person}{Clark Barrett}, \bibinfo{person}{Zhangyang Wang}, {and} \bibinfo{person}{Beidi Chen}.} \bibinfo{year}{2023}\natexlab{}.
\newblock \showarticletitle{{H2O}: Heavy-Hitter Oracle for Efficient Generative Inference of Large Language Models}.
\newblock \bibinfo{journal}{\emph{Advances in neural information processing systems}}  \bibinfo{volume}{36} (\bibinfo{year}{2023}), \bibinfo{pages}{34661--34710}.
\newblock


\bibitem[Zhao et~al\mbox{.}(2026c)]%
        {infllmv2}
\bibfield{author}{\bibinfo{person}{Weilin Zhao}, \bibinfo{person}{Zihan Zhou}, \bibinfo{person}{Zhou Su}, \bibinfo{person}{Chaojun Xiao}, \bibinfo{person}{Yuxuan Li}, \bibinfo{person}{Yanghao Li}, \bibinfo{person}{Yudi Zhang}, \bibinfo{person}{Weilun Zhao}, \bibinfo{person}{Zhen Li}, \bibinfo{person}{Yuxiang Huang}, \bibinfo{person}{Ao Sun}, \bibinfo{person}{Xu Han}, {and} \bibinfo{person}{Zhiyuan Liu}.} \bibinfo{year}{2026}\natexlab{c}.
\newblock \showarticletitle{{InfLLM-V2}: Dense-Sparse Switchable Attention for Seamless Short-to-Long Adaptation}. In \bibinfo{booktitle}{\emph{International Conference on Learning Representations}}, Vol.~\bibinfo{volume}{2026}. \bibinfo{pages}{69374--69390}.
\newblock


\bibitem[Zhao et~al\mbox{.}(2026b)]%
        {sparsespec}
\bibfield{author}{\bibinfo{person}{Yilong Zhao}, \bibinfo{person}{Jiaming Tang}, \bibinfo{person}{Kan Zhu}, \bibinfo{person}{Zihao Ye}, \bibinfo{person}{Chi-Chih Chang}, \bibinfo{person}{Chaofan Lin}, \bibinfo{person}{Jongseok Park}, \bibinfo{person}{Guangxuan Xiao}, \bibinfo{person}{Mohamed~S. Abdelfattah}, \bibinfo{person}{Mingyu Gao}, \bibinfo{person}{Baris Kasikci}, \bibinfo{person}{Song Han}, {and} \bibinfo{person}{Ion Stoica}.} \bibinfo{year}{2026}\natexlab{b}.
\newblock \showarticletitle{Accelerating Large-Scale Reasoning Model Inference with Sparse Self-Speculative Decoding}.
\newblock \bibinfo{journal}{\emph{Proceedings of Machine Learning and Systems}}  \bibinfo{volume}{8} (\bibinfo{year}{2026}), \bibinfo{pages}{251--265}.
\newblock


\bibitem[Zhao et~al\mbox{.}(2026a)]%
        {spin}
\bibfield{author}{\bibinfo{person}{Zihan Zhao}, \bibinfo{person}{Baotong Lu}, \bibinfo{person}{Shengjie Lin}, \bibinfo{person}{Yizou Chen}, \bibinfo{person}{Jing Liu}, \bibinfo{person}{Yanqi Zhang}, \bibinfo{person}{Ziming Miao}, \bibinfo{person}{Ming-Chang Yang}, \bibinfo{person}{Haiying Shen}, \bibinfo{person}{Qi Chen}, {and} \bibinfo{person}{Fan Yang}.} \bibinfo{year}{2026}\natexlab{a}.
\newblock \showarticletitle{Unifying Sparse Attention with Hierarchical Memory for Scalable Long-Context LLM Serving}.
\newblock \bibinfo{journal}{\emph{arXiv preprint arXiv:2604.26837}} (\bibinfo{year}{2026}).
\newblock


\bibitem[Zhong et~al\mbox{.}(2024)]%
        {distserve}
\bibfield{author}{\bibinfo{person}{Yinmin Zhong}, \bibinfo{person}{Shengyu Liu}, \bibinfo{person}{Junda Chen}, \bibinfo{person}{Jianbo Hu}, \bibinfo{person}{Yibo Zhu}, \bibinfo{person}{Xuanzhe Liu}, \bibinfo{person}{Xin Jin}, {and} \bibinfo{person}{Hao Zhang}.} \bibinfo{year}{2024}\natexlab{}.
\newblock \showarticletitle{{DistServe}: Disaggregating Prefill and Decoding for Goodput-Optimized Large Language Model Serving}. In \bibinfo{booktitle}{\emph{18th USENIX symposium on operating systems design and implementation (OSDI 24)}}. \bibinfo{pages}{193--210}.
\newblock


\end{thebibliography}

% \appendix
% \input{body/appendix}

\end{document}